\documentclass{aa}  

\usepackage{graphicx}
\usepackage{xcolor}
\usepackage{siunitx}

\usepackage{chemformula}
\usepackage{natbib}
\usepackage{multirow}
\usepackage{colortbl}
\usepackage{placeins}

\newcommand{\Msun}{\ensuremath{\mathrm{M}_\odot}\xspace}
\newcommand{\Teff}{\ensuremath{T_\mathrm{eff}}\xspace}
\newcommand{\numax}{\ensuremath{\nu_\mathrm{max}}\xspace}
\newcommand{\deltanu}{\ensuremath{\Delta\nu}\xspace}
\newcommand{\fov}{\ensuremath{f_\mathrm{ov}}\xspace}
\newcommand{\aMLT}{\ensuremath{\alpha_\mathrm{MLT}}\xspace}
\newcommand{\FeH}{\ensuremath{[\ch{Fe}/\ch{H}]}\xspace}
\newcommand{\aFe}{\ensuremath{[\alpha/\ch{Fe}]}\xspace}
\newcommand{\MH}{\ensuremath{[\ch{M}/\ch{H}]}\xspace}
\newcommand{\dYdZ}{\ensuremath{\Delta Y/\Delta Z}\xspace}

\makeatletter
\renewcommand*\aa@pageof{, page \thepage{} of \pageref*{LastPage}}
\makeatother

\usepackage{hyperref}
\hypersetup{colorlinks=true, allcolors=blue}

\begin{document} 

   \title{Inferring the efficiency of convective-envelope overshooting in Red Giant Branch stars}

   \subtitle{}

   \author{L. Briganti\inst{\ref{inst1}, \ref{inst2}}
    \and M. Tailo\inst{\ref{inst3}, \ref{inst1}}
    \and E. Ceccarelli\inst{\ref{inst2}, \ref{inst1}}
    \and A. Miglio\inst{\ref{inst1}, \ref{inst2}}
    \and M. Matteuzzi\inst{\ref{inst1}, \ref{inst2}}
    \and A. Mucciarelli\inst{\ref{inst1}, \ref{inst2}}
    \and A. Mazzi\inst{\ref{inst1}} 
    \and A. Bragaglia\inst{\ref{inst2}}
    \and S. Khan\inst{\ref{inst4}}
    }

   \institute{Department of Physics \& Astronomy "Augusto Righi", University of Bologna, via Gobetti 93/2, 40129 Bologna, Italy \\
            \email{lorenzo.briganti2@unibo.it}\label{inst1}
         \and
            INAF-Astrophysics and Space Science Observatory of Bologna, via Gobetti 93/3, 40129 Bologna, Italy\label{inst2} 
        \and
            INAF-Astronomic Observatory of Padova, vicolo dell'Osservatorio 5, 35122 Padova, Italy\label{inst3}
        \and 
            Institute of Physics, \'Ecole Polytechnique F\'ed\'erale de Lausanne (EPFL), Observatoire de Sauverny, 1290 Versoix, Switzerland\label{inst4}
             }

   \date{Received ; accepted }
 
  \abstract
   {The understanding of mixing processes in stars is crucial for improving our knowledge of the chemical abundances in stellar photospheres and of their variation with evolutionary phase. This is fundamental for many astrophysical issues on all scales, ranging from stellar evolution to the chemical composition, formation and evolution of stellar clusters and galaxies. \par
   
   Among these processes, convective-envelope overshooting is in dire need of a systematic calibration and comparison with predictions from multi-dimensional hydrodynamical simulations. The Red Giant Branch bump (RGBb) is an ideal calibrator of overshooting processes, since its luminosity depends on the maximum depth reached by the convective envelope after the first dredge-up. Indeed, a more efficient overshooting produces a discontinuity in the Hydrogen mass fraction profile deeper in the stellar interior and consequently a less luminous RGBb. \par
   
   In this work, we calibrated the overshooting efficiency by comparing the RGBb location predicted by stellar models with observations of stellar clusters with HST and \textit{Gaia} photometry, as well as solar-like oscillating giants in the \textit{Kepler} field. We explored  the metallicity range between $-2.02$ \si{dex} and $+0.35$ \si{dex} and found overshooting efficiencies ranging from $0.009^{+0.015}_{-0.016}$ to $0.062^{+0.017}_{-0.015}$. In particular, we found that the overshooting efficiency decreases linearly with $\MH$, with a slope of $(-0.010\pm0.006)\; \si{dex}^{-1}$. We suggest a possible explanation for this trend, linking it to the efficiency of turbulent entrainment at different metallicities.
}


   \maketitle

\section{Introduction}
\label{sec: Introduction}

Despite the fact that current stellar models are highly sophisticated and allow to simulate the life of stars from the Pre-Main Sequence (PMS) to their final evolution stages, there are still some open questions regarding their internal structure, such as the efficiency of mixing processes beyond convective boundaries. In particular, the understanding of convective-envelope overshooting in Red Giant Branch (RGB) stars could contribute to the explanation of photospheric abundances, e.g. leading to a more accurate calibration of masses (hence ages) of low-mass stars based on the $\ch{C}/\ch{N}$ ratio \citep{salaris2015, hasselquist2019, jofre2021}, or to a better understanding of Lithium depletion in solar-like stars \citep{baraffe2017, baraffe2021}. This phenomenon has also been extensively studied in the Sun, to better constrain its structure, by comparing the predictions of the models with helioseismic data \citep{christensen-dalsgaard2011, baraffe2022, zhang2022, buldgen2025}. \par 

Until now, overshooting has been included in 1D standard stellar models parametrically, shaping the border of the extended mixed region either with a step \citep[see, e.g.,][]{roxburgh1965, saslaw&schwarzschild1965, shaviv&salpeter1973, maeder1975, bressan1981, bressan1986} or an exponentially decaying (see \citealt{freytag1996, ventura1998, herwig2000} and Sec.~\ref{subsec: Data_Models}) function. 
On the other hand, in 2D \citep{freytag1996} and 3D \citep{meakin&arnett2007,blouin2023} hydrodynamical simulations the extension of the mixing region over the Schwarzschild border arises as a consequence of phenomena such as gravity wave generation and mass entrainment, which are not taken into account in the Mixing Length Theory \citep[MLT;][]{bohm-vitense1958, cox&giuli1968}, currently the most adopted theory to describe convective zones in stellar interiors. In particular, turbulent entrainment \citep{kantha1977, strang&fernando2001} describes how turbulent eddies diffuse into stable layers, increasing the size of the mixing zone, and could thus provide an explanation for the fact that a more efficient overshooting seems to be needed at lower metallicities (see \citealt{meakin&arnett2007, khan2021} and Sec.~\ref{subsec: Entrainment_law}). \par

A well-known calibrator for convective-envelope overshooting is the Red Giant Branch bump (RGBb). The RGBb is the direct consequence of a temporary drop in luminosity due to the encounter between the $\ch{H}$-burning shell, moving outwards, and the discontinuity in the $\ch{H}$ mass fraction left by the first dredge-up \citep{thomas1967, iben1968}. This makes the RGBb a tracer of the maximum depth reached by convection: in fact, a deeper convective envelope produces a deeper chemical discontinuity, making the RGBb to occur earlier along the RGB (i.e. at lower luminosities). \par

This feature was first empirically measured in the Globular Cluster (GC) NGC 104 by \cite{king1985} and then has been the main subject of many studies. In particular, by analysing a sample of 72 GCs, \cite{nataf2013} shows that the RGBb luminosity decreases with increasing metallicity \citep[as already observed by, e.g.,][]{fusipecci1990, cassisi&salaris1997, riello2003, dicecco2010}. Analogously \cite{khan2018} found that the RGBb frequency of maximum oscillation power (or, equivalently, the surface gravity, see \citealt{lindsay2022}) increases with increasing metallicity using field stars. However, it is well-established that standard stellar models overestimate the RGBb luminosity \citep{alongi1991, dicecco2010, joyce&chaboyer2015, fu2018} and that some envelope-overshooting is needed to match predictions and observations, with a more efficient overshooting at low metallicities \citep{dicecco2010, khan2018}. \par

Starting from this point, we want to extend the metallicity domain of these works and study the RGBb also in field stars through asteroseismology, as done in \cite{khan2018}, in order to calibrate the overshooting efficiency and investigate possible correlations of this quantity with stellar properties.  \par

The paper is organised in the following way: in Section~\ref{sec: Data} we present our grid of evolutionary tracks and our observational dataset, composed by stellar clusters and field stars; in Section~\ref{sec: Methods} we describe the procedure we adopted to determine the RGBb locations and to infer the overshooting efficiencies; in Section~\ref{sec: Results_and_discussion} we report and discuss our results. Finally, conclusions are drawn in Section~\ref{sec: Conclusions}.

\section{Models and data}
\label{sec: Data}

We analysed a grid of evolutionary tracks and two datasets, consisting in 17 Globular Clusters (GCs) and one Open Cluster (OC), and a sample of $\sim 3000$ RGB field stars from the catalogue by \cite{willett2025}: we will describe them in detail in the next sections.

\subsection{Evolutionary tracks}
\label{subsec: Data_Models}

We used \verb|MESA-r11701| \citep[Modules for Experiments in Stellar Astrophysics;][]{paxton2011, paxton2013, paxton2015, paxton2016, paxton2018, paxton2019} to compute a grid of evolutionary tracks with physical assumptions consistent with the ones adopted in Tailo et al. (in prep.). Specifically, we computed tracks for masses ranging from $0.8\; \Msun$ to $1.6\; \Msun$ (with steps of $0.2\; \Msun$), $\FeH$ from -2.1 \si{dex} to +0.3 \si{dex} (with steps of 0.2 \si{dex}), $\aFe$ from +0.0 \si{dex} to +0.4 \si{dex} (with steps of 0.2 \si{dex}), $\aMLT = 2.090, 2.290$ (the former corresponding to the value that gives a better fit to APOGEE effective temperatures, as explained in Sec.~\ref{subsec: mixing_length}, the latter resulting from the solar calibration). We employed a linear enrichment law with $\dYdZ = 1.245$ and $Y_P = 0.2485$ \citep{komatsu2011}. We included atomic diffusion and used atmosphere boundary conditions from \cite{krishnaswamy1966}. All the evolutionary tracks were computed from the PMS to the $\ch{He}$-flash. \par

In addition, we included, at the base of the convective envelope, diffusive overshooting \citep{freytag1996, herwig2000}, i.e. an extra-mixing process in which the diffusion coefficient decays exponentially with the distance from the convective boundary defined using the Schwarzschild criterion (located at distance $r_\mathrm{CE}$ from the stellar centre). Since in \verb|MESA| the diffusion coefficient in the convective zone goes to zero at $r_\mathrm{CE}$, the switch between convection and overshooting is set to occur at $r_{0} = r_\mathrm{CE}+f_{0}H_\mathrm{P,CE}$, where $f_{0}$ is a parameter that regulates the diffusion coefficient in the convective envelope at $r=r_{0}$ and $H_\mathrm{P,CE}$ is the pressure scale height at the base of the convective envelope. The diffusion coefficient in the overshooting zone is therefore written as:
\begin{equation}
    D_\mathrm{ov} = D_{0}\exp{\biggl(-\frac{2(r_{0}-r)}{\fov H_\mathrm{P,CE}}\biggr)} \qquad (r<r_{0}),
\end{equation}
where $D_\mathrm{ov}$ and $D_{0}$ are the diffusion coefficients at distance $r$ and $r_{0}$ from the stellar centre, and $\fov$ is the overshooting efficiency.
In particular, we set $f_{0} = 0.001$ and varied the overshooting efficiency $\fov$ from 0.000 to 0.125 (with steps of 0.025), obtaining a grid of 2340 evolutionary tracks (1170 for each value of $\aMLT$).

\subsection{Stellar clusters}
\label{subsec: Data_Clusters}

\renewcommand{\arraystretch}{1.4}
\begin{table*}
\caption{Properties (age, chemical composition and reddening) and RGBb luminosities for all the clusters in our sample.}            
\label{tab:Clusters_properties}     
\centering                          
\begin{tabular}{lrccccc}        
\hline \hline          
NGC & $t_\mathrm{Age}$ [Gyr] & $\FeH$ & $\aFe$ & $\MH$ & E(B-V) & $\log L_\mathrm{RGBb}\; [L_\odot]$   \\    
\hline                       
    1261 & $10.83 \pm 0.42$ & $-1.27 \pm 0.08$ & $0.360 \pm 0.040$ (1) & $-1.01 \pm 0.09$ & 0.01 (P16) & $1.81^{+0.01}_{-0.01}$ \\
    2298 & $12.84 \pm 0.59$ & $-1.96 \pm 0.04$ & $0.500 \pm 0.100$ (2) & $-1.58 \pm 0.09$ & 0.22 (P16) & $1.85^{+0.02}_{-0.02}$ \\
    4590 & $12.17 \pm 0.51$ & $-2.27 \pm 0.04$ & $0.350 \pm 0.100$ (2) & $-2.02 \pm 0.09$ & 0.06 (P16) & $2.07^{+0.01}_{-0.01}$ \\
    5053 & $12.68 \pm 0.47$ & $-2.30 \pm 0.08$ & $0.340 \pm 0.070$ (3) & $-2.06 \pm 0.10$ & 0.01 (P16) & --- \\
    5466 & $13.02 \pm 0.48$ & $-2.31 \pm 0.09$ & $0.330 \pm 0.100$ (2) & $-2.07 \pm 0.12$ & 0.02 (P16) & $2.11^{+0.02}_{-0.02}$ \\
    6101 & $12.60 \pm 0.53$ & $-1.98 \pm 0.07$ & $0.400 \pm 0.100$ (4) & $-1.69 \pm 0.11$ & 0.13 (P16) & $2.00^{+0.01}_{-0.01}$ \\
    6121 & $12.18 \pm 0.45$ & $-1.18 \pm 0.02$ & $0.510 \pm 0.100$ (2) & $-0.80 \pm 0.09$ & 0.44 (G19) & $1.69^{+0.01}_{-0.01}$ \\
    6144 & $13.36 \pm 0.54$ & $-1.82 \pm 0.05$ & $0.400 \pm 0.100$ (4) & $-1.53 \pm 0.10$ & 0.41 (G19) & $1.88^{+0.02}_{-0.02}$ \\
    6218 & $12.97 \pm 0.47$ & $-1.33 \pm 0.02$ & $0.410 \pm 0.100$ (2) & $-1.03 \pm 0.08$ & 0.19 (P16) & $1.77^{+0.01}_{-0.01}$ \\
    6362 & $12.86 \pm 0.44$ & $-1.07 \pm 0.05$ & $0.370 \pm 0.120$ (5) & $-0.80 \pm 0.11$ & 0.07 (P16) & $1.68^{+0.01}_{-0.01}$ \\
    6397 & $13.06 \pm 0.42$ & $-1.99 \pm 0.02$ & $0.360 \pm 0.100$ (2) & $-1.73 \pm 0.08$ & 0.19 (P16) & $2.00^{+0.01}_{-0.01}$ \\
    6535 & $12.17 \pm 0.63$ & $-1.79 \pm 0.07$ & $0.320 \pm 0.110$ (6) & $-1.56 \pm 0.11$ & 0.42 (G19) & --- \\
    6584 & $11.75 \pm 0.47$ & $-1.50 \pm 0.09$ & $0.410 \pm 0.080$ (7) & $-1.20 \pm 0.11$ & 0.10 (P16) & $1.85^{+0.01}_{-0.01}$ \\
    6637 & $12.19 \pm 0.53$ & $-0.59 \pm 0.07$ & $0.310 \pm 0.100$ (2) & $-0.37 \pm 0.10$ & 0.15 (P16) & $1.55^{+0.01}_{-0.01}$ \\
    6652 & $12.48 \pm 0.46$ & $-0.76 \pm 0.14$ & $0.200 \pm 0.200$ (8) & $-0.62 \pm 0.20$ & 0.12 (P16) & $1.57^{+0.02}_{-0.02}$ \\
    6717 & $12.89 \pm 0.50$ & $-1.26 \pm 0.07$ & $0.400 \pm 0.100$ (4) & $-0.97 \pm 0.11$ & 0.26 (G19) & $1.77^{+0.02}_{-0.02}$ \\
    6934 & $11.63 \pm 0.46$ & $-1.56 \pm 0.09$ & $0.400 \pm 0.100$ (1) & $-1.27 \pm 0.12$ & 0.10 (G19) & $1.88^{+0.01}_{-0.01}$ \\
    6791 & $8.30 \pm 0.30$ & $+0.29 \pm 0.09$ & $0.092 \pm 0.007$ (9) & $+0.35 \pm 0.09$ & 0.14 (B12) & $1.37^{+0.04}_{-0.04}$ \\
\hline
\end{tabular}
\tablefoot{The references for age and $\FeH$ are \cite{kruijssen2019} and \cite{carretta2010}, respectively, for all the clusters but NGC 6791, for which we took these quantities from \cite{brogaard2012} and \cite{brogaard2011}, respectively. References for $\aFe$ are listed below, $\MH$ is computed with equation (3) from \cite{salaris1993}. The references for the reddening values are labelled as follows: P16 for \cite{planck2016}, G19 for \cite{green2019} and B12 for \cite{brogaard2012}. Finally, the RGBb luminosities have been derived with the methods described in Sec.~\ref{subsec: Methods_clusters} (for NGC 5053 and NGC 6535 no RGBb luminosity is provided, since there are not enough stars on the RGB). The errors on $\log L_\mathrm{RGBb}$ include the random error from the fit, the systematic due the number of stars (see Appendix~\ref{appendix: Script_testing}) and the error on the distance modulus \citep{baumgardt2018}.}
\tablebib{(1) \cite{marino2021}; (2) \cite{carretta2010}; (3) \cite{sbordone2015}; (4) \cite{tailo2020}; (5) \cite{massari2017}; (6) \cite{bragaglia2017}; (7) \cite{omalley&chaboyer2018}; (8) \cite{sharina&shimansky2020}; (9) \cite{linden2017}}
\end{table*}
\renewcommand{\arraystretch}{1}

As mentioned at the beginning of this section, our dataset includes 17 GCs. Since the $\ch{He}$ content affects the position of the RGBb (see \citealt{cassisi&salaris1997, salaris2006} and Appendix~\ref{app: Y_dY/dZ}), we selected GCs with $\delta Y_\mathrm{2G, 1G}^\mathrm{avg} \leq 0.01$ and $\delta Y_\mathrm{2G, 1G}^\mathrm{max} \leq 0.02$, where $\delta Y_\mathrm{2G, 1G}^\mathrm{avg}$ and $\delta Y_\mathrm{2G, 1G}^\mathrm{max}$ are, respectively, the average and the maximum $\ch{He}$ mass fraction differences between the second (2G) and the first (1G) generation of stars \citep[see][]{milone2018}. \par

We used both HST photometry collected in the HUGS \citep[HST UV Globular cluster Survey;][]{piotto2015, nardiello2018} catalogue and \textit{Gaia} EDR3 photometry \citep{gaia2016, gaia2023}, collected in the catalogue from \cite{vasiliev&baumgardt2021}. Both have been corrected for differential reddening, as explained in Sec.~\ref{subsec: Methods_clusters}. A probability membership based on proper motions is provided for all sources of each cluster both in the HUGS survey and in the \cite{vasiliev&baumgardt2021} catalogue. Consequently, to avoid contamination by field stars, we limited our analysis only to stars with membership probability $\geq 90\%$. \par

Age and $\FeH$ of each GC were taken from \cite{kruijssen2019} and \cite{carretta2010}, respectively, while for $\alpha$ elements abundance, when a spectroscopic value was not available in \cite{carretta2010}, we used literature values as indicated in Table \ref{tab:Clusters_properties}. The only exceptions are NGC 6101, NGC 6144 and NGC 6717, for which we adopted $\aFe = 0.4$ \si{dex}, as suggested in \cite{dotter2010, tailo2020}, since spectroscopic measurements are not available in literature for these GCs.

For the reddening, we used the values from the 3D extinction map by \cite{green2019}, based on Pan-STARRS~1 photometry \citep{panstarrsChambers2016}. Because the map is limited to a declination of $-30$\textdegree, for the remaining part of the sky we used extinction values from the \cite{planck2016} 2D extinction map. The actual querying of the maps was done with the \texttt{dustmaps}\footnote{\url{https://github.com/gregreen/dustmaps}} Python module \citep{dustmaps2018}. \par

In order to explore also super-solar metallicities, we included in the dataset NGC 6791, an open cluster for which \textit{Gaia} DR3 photometry is collected in the catalogue from \cite{hunt&reffert2023}. For this cluster, we selected stars with membership probability $\geq 70\%$ and adopted the age and the reddening reported in \cite{brogaard2012}, while $\FeH$ and $\aFe$ were taken from \cite{brogaard2011} and \cite{linden2017}, respectively. All the properties of the stellar clusters in our sample are reported in Table \ref{tab:Clusters_properties}.

\subsection{Field stars}
\label{subsec: Data_Field_stars}
We started from a sample of nearly $10^4$ RGs listed in the catalogue from \cite{willett2025}. This catalogue contains asteroseismic data from \textit{Kepler} \citep{borucki2010, gilliland2018}, astrometry from \textit{Gaia} DR3 \citep{gaia2016, gaia2023} and spectroscopy from APOGEE DR17 \citep{majewski2017, abdurrouf2022}. In addition, it includes stellar masses computed from the asteroseismic observables $\numax$ (frequency of maximum oscillation power) and $\deltanu$ (large frequency separation), using the Bayesian tool \verb|PARAM| \citep{dasilva2006, rodrigues2014, rodrigues2017, miglio2021a}. The distinction of RGs undergoing shell-hydrogen-burning (hence, RGB stars) and core-helium-burning (i.e. Red-Clump stars) was done using the flag contained in \cite{yu2018}, based on the distribution of oscillation amplitude, granulation power and width of power excess. \par

We restricted the RGB sample to stars with masses lower than $1.5\; \Msun$, to avoid effects from the convective core overshooting \citep[see e.g.][]{khan2018} and from rotational mixing \citep[see e.g.][]{charbonnel&lagarde2010, lagarde2012, beyer&white2024} during the Main Sequence (MS) phase. Furthermore, we selected only stars with good astrometric quality (\verb|RUWE| $< 1.4$, following prescriptions from \citealt{lindegren2018}) and parallaxes with errors up to $10\%$, according to \textit{Gaia} DR3, and considered stars with available luminosity derived from the parallaxes corrected for zero-point offset according to the scheme of \cite{lindegren2021}. In the end, the final sample is composed of 2737 RGB stars. \par

\section{Methods}
\label{sec: Methods}
\begin{figure}
    \centering
    \includegraphics[width=\columnwidth]{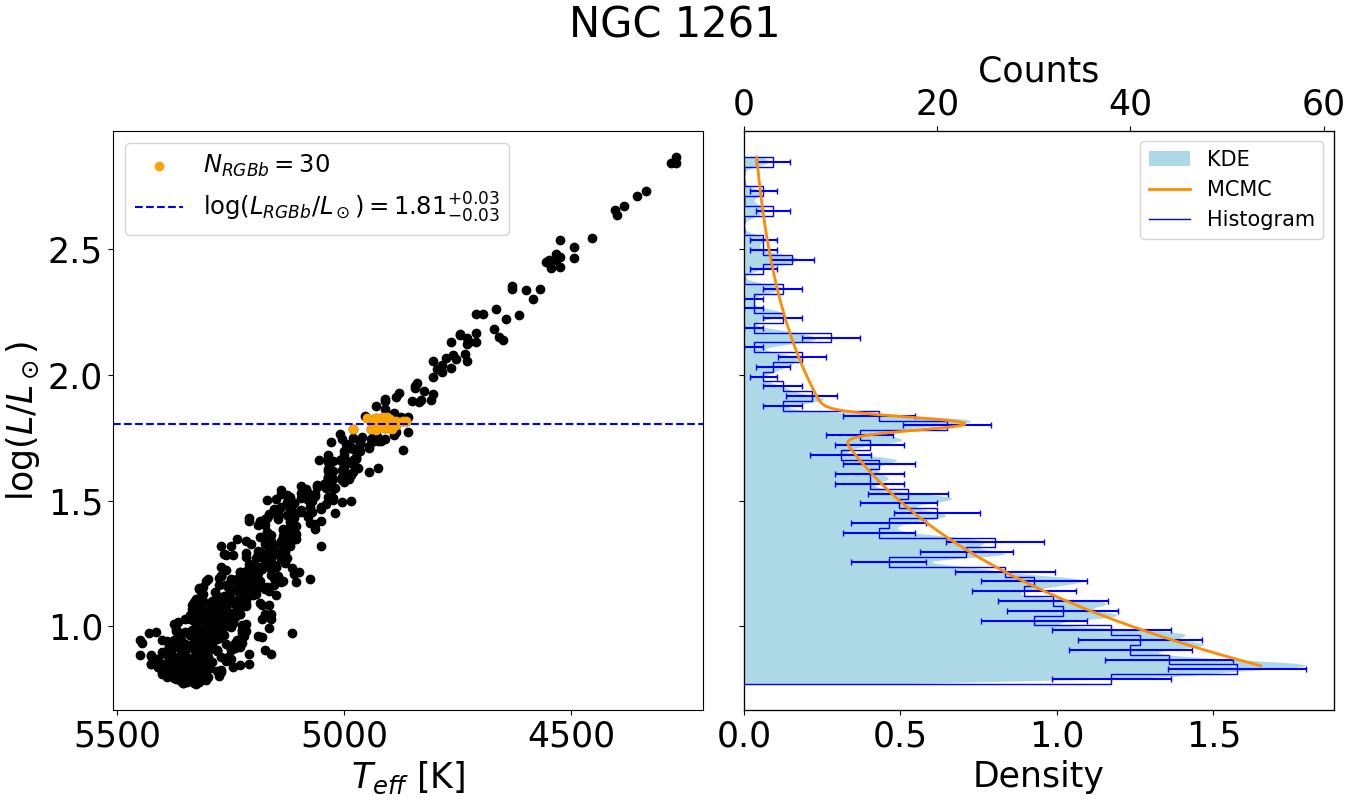}
    \caption{\textit{Left panel}: HRD of the RGB of NGC 1261, with RGBb stars highlighted in orange and a blue dashed line that refers to the RGBb luminosity obtained from the fit. The errors on $\log L_\mathrm{RGBb}$ already include the systematics due to the number of RGB stars (see Appendix~\ref{appendix: Script_testing}) and the distance modulus \citep{baumgardt2018}. \textit{Right panel}: KDE (light blue coloured zone) of the luminosity distribution (lower labels) and relative fit (orange line). The blue line represents the histogram of the RGB luminosity with Poissonian errors (upper labels).}
    \label{fig: 1261_fit}
\end{figure}

A consistent procedure was applied to both simulated and observational data, involving the selection of RGB stars and the computation of a Kernel Density Estimation (KDE) of their distribution in $L$ and $\numax$ (where available) using the Python package \verb|SciPy| \citep{virtanen2020}. Subsequently, the KDE was fitted using a Markov chain Monte Carlo method, implemented through the Python package \verb|emcee| \citep{foreman-mackey2013}, to accurately determine the location of the RGBb. \par

We adopted a fitting function consisting of the superposition of an exponential background and a Gaussian function:
\begin{equation}
	f (x;\lambda, \mu, \sigma, A, B) = Ae^{\lambda x} + 
		Be^{-\frac{(x-\mu)^2}{2\sigma^2}},
	\label{eq: Fit_function}
\end{equation}
with $x$ being $\log L$ (or $\log\numax$), $\lambda$ the slope of the exponential background (negative for $x = \log L$ and positive for $x = \log\numax$), $\mu$ and $\sigma$ the mean and standard deviation of the Gaussian function, and $A$ and $B$ the normalisation factors. We considered as RGBb stars all the stars with $\log L$ ($\log\numax$) between $\mu-\sigma$ and $\mu+\sigma$. An example of the fitting procedure for the GC NGC 1261 is reported in Fig.~\ref{fig: 1261_fit}. \par

Finally, for both the field stars and stellar clusters, we  inferred the overshooting efficiency by comparing our observational and theoretical results. In the following sections we will describe in detail these procedures for the evolutionary tracks and the two observational datasets. \par

\subsection{Simulated data}
\label{subsec: Methods_Models}
In the case of evolutionary tracks, RGB stars were defined as points cooler than the MS Turn-Off and with a surface gravity lower than a threshold value (in order to avoid the Sub-Giant Branch). \par

After the selection of the RGB, we generated for each track a synthetic population extracting 5000 stars uniformly distributed in age and we computed through interpolation the luminosity ($\log L$), the effective temperature ($\Teff$) and the frequency of maximum oscillation power ($\numax$), adding to these three quantities normally distributed deviations to simulate photometric errors on magnitudes and colours. After that we computed the KDE both for $\log L$ and $\log\numax$, and fitted it with equation~\ref{eq: Fit_function}. The fitting of simulations was used also to test the script and compute the systematics on the estimated RGBb location due to the number of stars in the fitted sample, as described in Appendix~\ref{appendix: Script_testing}.

\subsection{Analysis of stellar clusters}
\label{subsec: Methods_clusters}
 
   \begin{figure}
   \centering
   \includegraphics[width=.49\columnwidth]{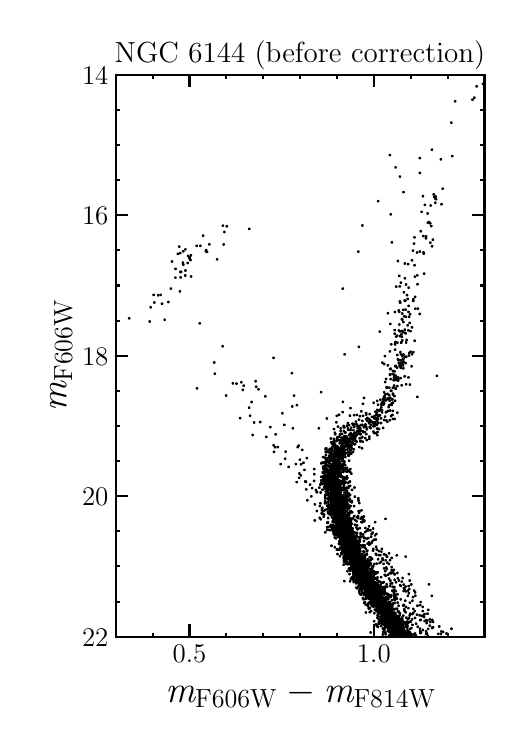} 
   \includegraphics[width=.49\columnwidth]{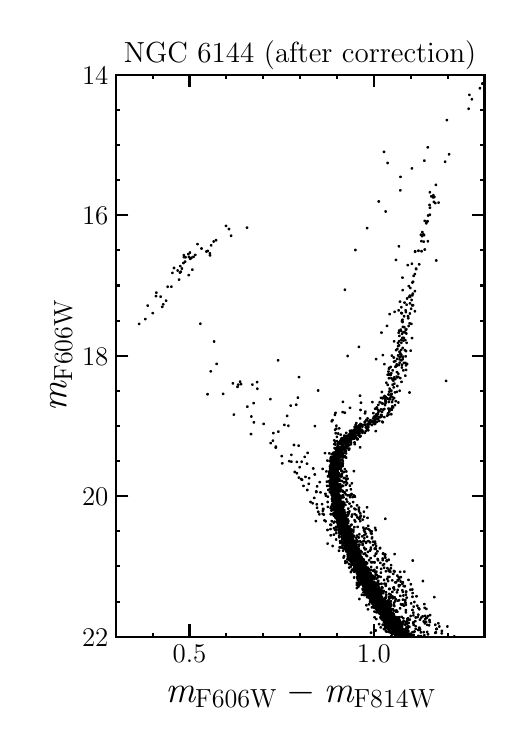}    
   \caption{Comparison between the CMDs of NGC 6144 before (left panel) and after (right panel) applying the differential reddening correction.}
              \label{fig:DR_correction}
    \end{figure} 
  
   \begin{figure}
   \centering
   \includegraphics[width=\columnwidth]{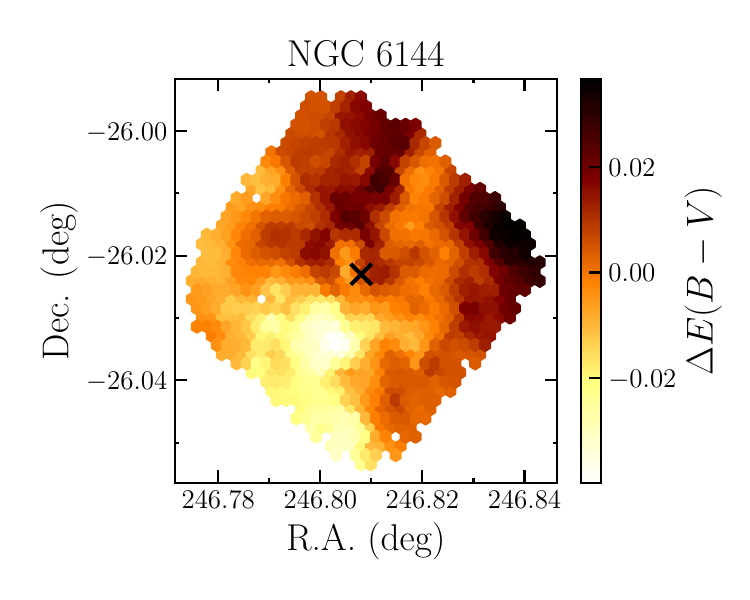} 
   \caption{Differential reddening map for the globular cluster NGC 6144 obtained using HST data sampling a radial range up to $\sim 1.0$ half-light radius\protect\footnotemark. The black cross marks the centre of the cluster \citep{vasiliev&baumgardt2021}.}
    \label{fig:DR_map}
    \end{figure} 
\footnotetext{\url{https://people.smp.uq.edu.au/HolgerBaumgardt/globular/parameter}}

For each GC, before fitting the RGB luminosity distribution and determining the overshooting efficiency, we determined the average mass of stars on the RGB and de-reddened the data, in order to compute the effective temperature and the luminosity of each star. All these procedures are described in detail in the following paragraphs. \par

We derived from the models a mass-age-metallicity scaling relation, which we used to infer the average stellar mass at the 
RGBb ($M_\mathrm{RGBb}$) for each cluster, given the age and the chemical composition:
\begin{equation}
    \biggl(\frac{\mathrm{M}_\mathrm{RGBb}}{\Msun}\biggr) = A\biggl(\frac{t_\mathrm{Age}}{10^9\; \si{yr}}\biggr)^\alpha + B\MH^\beta,
    \label{eq: mass-age-mh_relation}
\end{equation}
with $A = 1.906 \pm 0.003$, $\alpha = (-2.671 \pm 0.003)\cdot10^{-1}$, $B = (1.551 \pm 0.007)\cdot10^{-1}$ and $\beta = (5.665 \pm 0.013)\cdot10^{-1}$. This scaling relation has been computed at the RGBb luminosity for each model. The coefficients $A$, $\alpha$, $B$ and $\beta$ show very little variation with the overshooting efficiency (see Appendix~\ref{appendix: scaling_relation}), hence we decided to work with the median value of each parameter. \par

In order to perform the fit on the best possible data, after the membership selection, we corrected the HST and \textit{Gaia} photometry of 16 and 7 GCs respectively for the effects of differential reddening following the method described in \cite{milone2012}. To do so, we focused our attention on RGB stars to define the reference sample. Even though they are less numerous, we prefer RGB over MS stars since their photometric error is lower. For each star in the target globular clusters, the differential reddening value, $\Delta E(B - V)$, was computed as the median offset from the RGB mean ridge line, defined in a Colour-Magnitude Diagram (CMD) tilted along the reddening vector, among the 30 - 60 closest reference stars. We assumed the \cite{cardelli1989} extinction law, adopting $R_\mathrm{V} = 3.1$. This iterative procedure was repeated for each star until the residual on the $\Delta E(B - V)$ matched the photometric error. Typically, we reach convergence after two iterations. In Figure \ref{fig:DR_correction} we show an example of the differential reddening on the CMD of NGC 6144, while in Figure \ref{fig:DR_map} we plot the reddening map we derive. \par

After the de-reddening procedure, we converted colours and magnitudes into effective temperatures and luminosities, respectively: this was done using the bolometric corrections (BCs) from \cite{casagrande2014}, the distances from the 2023 version of the \cite{baumgardt2018} catalogue and the reddenings reported in Table~\ref{tab:Clusters_properties}. \par

For each GC, we built a synthetic grid of stars with different values of surface gravity and effective temperature ($0.2 \le \log g \le 5.5$, with steps of 0.1, $3000\; \si{K} \le \Teff \le 8000\; \si{K}$ with steps of $5\; \si{K}$) and computed for each point the BCs in the F606W and F814W bands. We then inferred $\log g$ for each star from the isochrone relative to the cluster, and assigned the temperature in the grid corresponding to the value of $BC_{F814W} - BC_{F606W}$ closest to the real colour of the star. We compared our evolutionary tracks with the corresponding ones from the PARSEC database \citep{bressan2012, chen2014, chen2015, fu2018}: given the small difference in $\log g$ at fixed luminosity ($\langle\Delta\log g_{\texttt{MESA}-\mathrm{PARSEC}}\rangle \sim -0.02$), we decided to use PARSEC isochrones instead of computing our own ones. \par

After obtaining the Hertzsprung-Russell Diagram (HRD) for each cluster, we used \verb|TOPCAT| \citep[Tool for OPerations on Catalogs And Tables;][]{taylor2005} to properly separate the RGB from the Horizontal Branch and the Asymptotic Giant Branch, and fitted the KDE of the luminosity distribution with equation~\ref{eq: Fit_function}. We performed the fit for luminosities greater than the one corresponding to the KDE maximum, since the cut at the base of the RGB affects the counts in the lowest luminosity bins and, hence, the slope of the exponential background. We excluded NGC 5053 and NGC 6535 because their RGBs contain too few stars to make a robust prediction on the RGBb position (see App.~\ref{appendix: Script_testing}). The HRDs of 7 out of 17 GCs were completed adding stars from the \cite{vasiliev&baumgardt2021} catalogue, containing \textit{Gaia} DR3 photometry \citep{gaia2016, gaia2023}: for these stars we used the colour-temperature relation from \cite{mucciarelli2021} and the BCs from \cite{andrae2018}. \par

\begin{figure}
    \centering
    \includegraphics[width=\columnwidth]{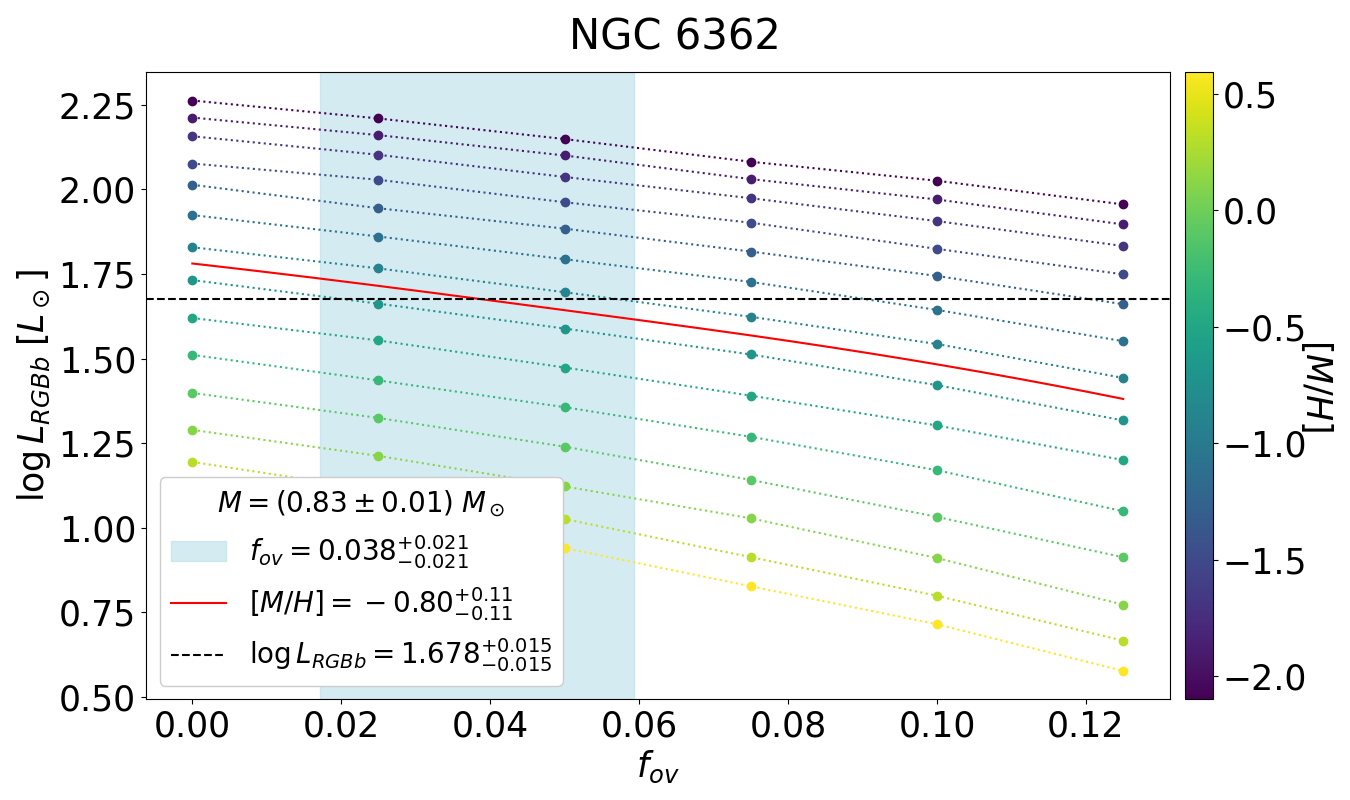}
    \caption{Representation, in the $\fov-\log L$ plane, of the interpolation procedure to derive the overshooting efficiency for the GC NGC 6362. The expectation values of $\MH$ and $\log L_{RGBb}$ are represented by the red solid line and the black dashed one, respectively, and are plotted on the grid of models computed at the cluster mass (coloured by metallicity). The light blue shaded zone represents the confidence interval at 68\% of the derived overshooting efficiency.}
    \label{fig: 6362_interpolation}
\end{figure}

Once the luminosity of the RGBb was determined, we restricted our sample to clusters with at least 10 RGBb stars (ending up with 11 GCs out of 17) and we performed an interpolation in the space $M - \log L - \MH - \fov$ to derive the overshooting efficiency for each GC. This operation was repeated $10^4$ times extracting randomly a pair ($\MH$, $\log L$) from a normal distribution centred on the expected values of the two quantities and with $\sigma$'s corresponding to the associated errors. An example is provided in Fig.~\ref{fig: 6362_interpolation}. \par

For the OC NGC 6791 the methods were the same, except for the de-reddening procedure, that we did not perform due to a low number of stars along the RGB, and the conversion of HST colours and magnitudes into temperatures and luminosities, since for this cluster we used only \textit{Gaia} photometry. Furthermore, for this cluster we used the distance modulus reported in \cite{brogaard2011}.

\subsection{Analysis of field stars}
\label{subsec: Methods_Field_stars}
\renewcommand{\arraystretch}{1.4}
\begin{table*}
\caption{Properties and RGBb locations for the bins of field stars in our sample.}            
\label{tab: field_stars_bins}     
\centering                          
\begin{tabular}{c c c c c c c c c}        
\hline \hline              
$\mathrm{M}$ range $[\Msun]$ & $\mathrm{M}\; [\Msun]$ & $\MH$ range & $\MH$ & $\FeH$ & $\aFe$ & $N_\star$ & $\log L_\mathrm{RGBb}$ & $\log\nu_\mathrm{max, RGBb}$ \\
\hline                  
    $[0.9, 1.1[$ & $0.97^{+0.07}_{-0.04}$ & $[-0.6, -0.4[$ & $-0.47^{+0.04}_{-0.05}$ & $-0.64^{+0.10}_{-0.08}$ & $+0.25^{+0.04}_{-0.13}$ & $61$ & $1.60^{+0.02}_{-0.02}$ & $0.97^{+0.02}_{-0.02}$\\
    $[0.9, 1.1[$ & $1.01^{+0.06}_{-0.06}$ & $[-0.4, -0.2[$ & $-0.29^{+0.06}_{-0.06}$ & $-0.41^{+0.09}_{-0.10}$ & $+0.20^{+0.06}_{-0.11}$ & $206$ & $1.53^{+0.01}_{-0.01}$ & $1.01^{+0.01}_{-0.01}$\\
    $[0.9, 1.1[$ & $1.02^{+0.05}_{-0.05}$ & $[-0.2, +0.0[$ & $-0.11^{+0.07}_{-0.06}$ & $-0.19^{+0.09}_{-0.09}$ & $+0.12^{+0.10}_{-0.06}$ & $263$ & $1.46^{+0.01}_{-0.01}$ & $1.02^{+0.01}_{-0.01}$\\
    $[0.9, 1.1[$ & $1.04^{+0.04}_{-0.08}$ & $[+0.0, +0.2[$ & $+0.07^{+0.07}_{-0.05}$ & $+0.04^{+0.08}_{-0.07}$ & $+0.05^{+0.07}_{-0.03}$ & $146$ & $1.46^{+0.01}_{-0.01}$ & $1.04^{+0.01}_{-0.01}$\\
    $[1.1, 1.3[$ & $1.17^{+0.07}_{-0.05}$ & $[-0.4, -0.2[$ & $-0.26^{+0.05}_{-0.07}$ & $-0.33^{+0.07}_{-0.08}$ & $+0.08^{+0.02}_{-0.02}$ & $140$ & $1.62^{+0.01}_{-0.01}$ & $1.17^{+0.02}_{-0.02}$\\
    $[1.1, 1.3[$ & $1.20^{+0.06}_{-0.06}$ & $[-0.2, +0.0[$ & $-0.10^{+0.07}_{-0.07}$ & $-0.14^{+0.08}_{-0.08}$ & $+0.05^{+0.03}_{-0.02}$ & $259$ & $1.55^{+0.01}_{-0.01}$ & $1.20^{+0.01}_{-0.01}$\\
    $[1.1, 1.3[$ & $1.19^{+0.07}_{-0.06}$ & $[+0.0, +0.2[$ & $+0.08^{+0.07}_{-0.06}$ & $+0.06^{+0.07}_{-0.07}$ & $+0.03^{+0.02}_{-0.02}$ & $276$ & $1.47^{+0.01}_{-0.01}$ & $1.19^{+0.01}_{-0.01}$\\
    $[1.1, 1.3[$ & $1.18^{+0.06}_{-0.05}$ & $[+0.2, +0.4[$ & $+0.27^{+0.07}_{-0.05}$ & $+0.26^{+0.07}_{-0.05}$ & $+0.02^{+0.01}_{-0.02}$ & $144$ & $1.43^{+0.01}_{-0.01}$ & $1.18^{+0.01}_{-0.01}$\\
    $[1.3, 1.5[$ & $1.38^{+0.08}_{-0.06}$ & $[-0.4, -0.2[$ & $-0.26^{+0.04}_{-0.07}$ & $-0.32^{+0.06}_{-0.08}$ & $+0.08^{+0.03}_{-0.03}$ & $89$ & $1.62^{+0.02}_{-0.02}$ & $1.38^{+0.02}_{-0.02}$\\
    $[1.3, 1.5[$ & $1.39^{+0.06}_{-0.07}$ & $[-0.2, +0.0[$ & $-0.09^{+0.06}_{-0.07}$ & $-0.12^{+0.08}_{-0.08}$ & $+0.04^{+0.02}_{-0.02}$ & $184$ & $1.66^{+0.01}_{-0.01}$ & $1.39^{+0.01}_{-0.01}$\\
    $[1.3, 1.5[$ & $1.38^{+0.07}_{-0.06}$ & $[+0.0, +0.2[$ & $+0.08^{+0.07}_{-0.06}$ & $+0.08^{+0.07}_{-0.06}$ & $+0.02^{+0.02}_{-0.01}$ & $156$ & $1.62^{+0.01}_{-0.01}$ & $1.38^{+0.01}_{-0.01}$\\
\hline
\end{tabular}
\tablefoot{Columns are: range and median value of stellar mass (in solar masses), range and median value of $\MH$, median values of $\FeH$ and $\aFe$, number of stars in the bin, and RGBb locations in $\log L$ (expressed in $L_\odot$) and $\log\numax$ (expressed in $\mu\mathrm{Hz}$).}
\end{table*}
\renewcommand{\arraystretch}{1}

We divided our 2737 RGB stars in bins of different masses ([0.7, 0.9[, [0.9, 1.1[, [1.1, 1.3[, [1.3, 1.5[, in solar masses) and metallicities ([-0.8, -0.6[, [-0.6, -0.4[, [-0.4, -0.2[, [-0.2, +0.0[, [+0.0, +0.2[, [+0.2, +0.4[ \si{dex}). We considered only bins with at least 60 stars, ending up 1924 stars arranged in 11 bins (see Tab.~\ref{tab: field_stars_bins}). \par

Unlike what we have done for synthetic populations and stellar clusters, for each bin, we fitted only the highest peak with a gaussian function, both in the $\log L$ axis and $\log\numax$ axis, since the exponential background was not well identifiable in all the combinations of mass and metallicity bins. After that we performed an interpolation in the space $M - \log L\; (\log\numax) - \MH - \fov$ to derive the overshooting efficiency for each bin, in the same way we have done for stellar clusters (see Sec.~\ref{subsec: Methods_clusters}). \par

\section{Results and discussion}
\label{sec: Results_and_discussion}

\begin{figure}
    \centering
    \includegraphics[width=\columnwidth]{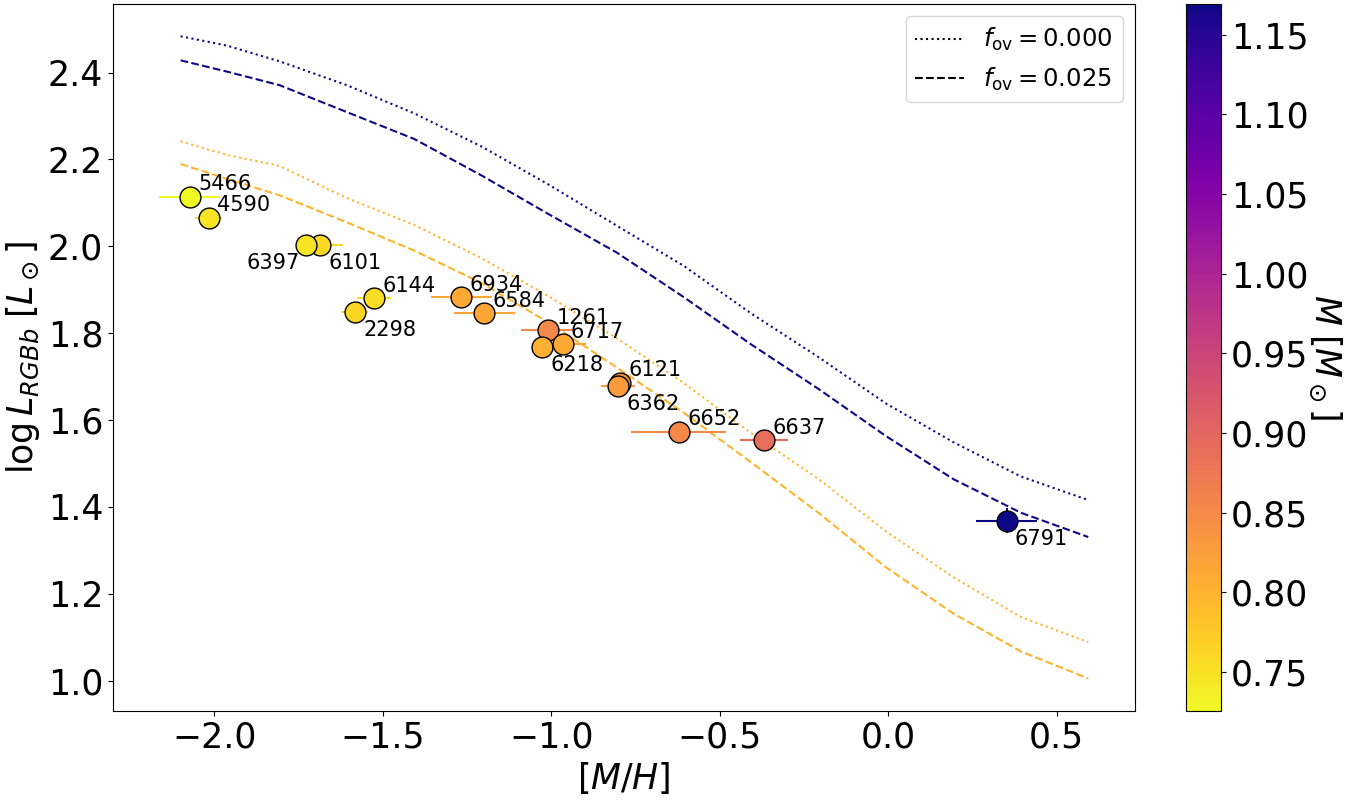}
    \caption{RGBb luminosity as a function of $\MH$ for the stellar clusters in our sample (coloured by $M_\mathrm{RGBb}$). The RGBb luminosity decreases with increasing $\MH$ and decreasing stellar mass (see e.g., NGC 1261 and NGC 6218). The lines correspond to the RGBb luminosities, at different metallicities, of evolutionary tracks with $\mathrm{M} = (0.8, 1.17)\Msun$ (orange, blue) and $\fov = 0.000, 0.025$ (dotted, dashed).}
    \label{fig: logL_MH_GC}
\end{figure}

\begin{figure*}
    \centering
    \includegraphics[width=17cm]{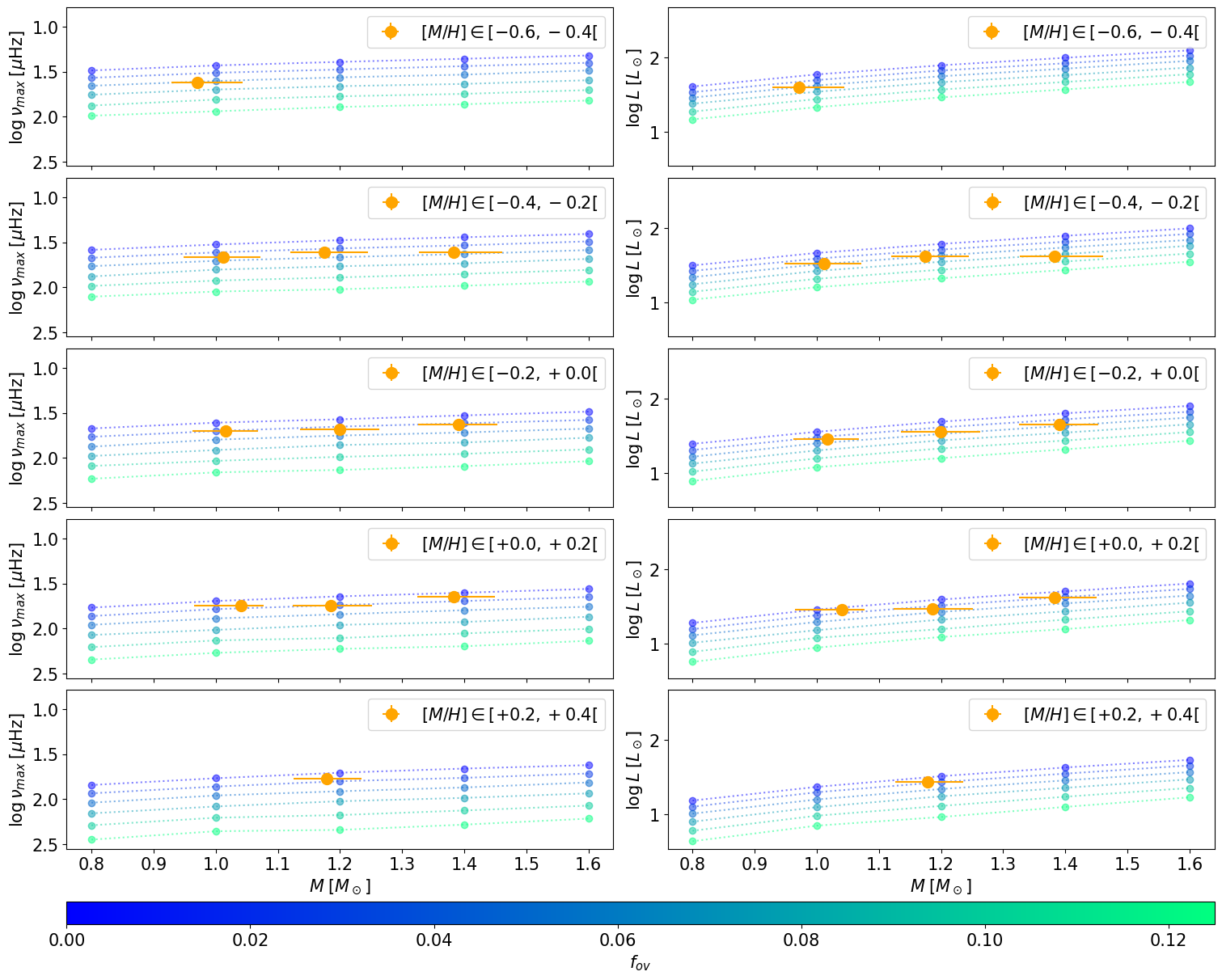}
    \caption{Location of the RGBb in $\log\numax$ (left) and $\log L$ (right), as a function of stellar mass, for \textit{Kepler} stars divided in metallicity bins (orange points). The points coloured smoothly from green to blue represent the location of the RGBb in the same metallicity bins for different overshooting efficiencies. The colour-coding for $f_\mathrm{ov}$ is reported in the bottom horizontal bar.}
    \label{fig: Bump_trends_Mass}
\end{figure*}

\begin{figure*}
    \centering
    \includegraphics[width=17cm]{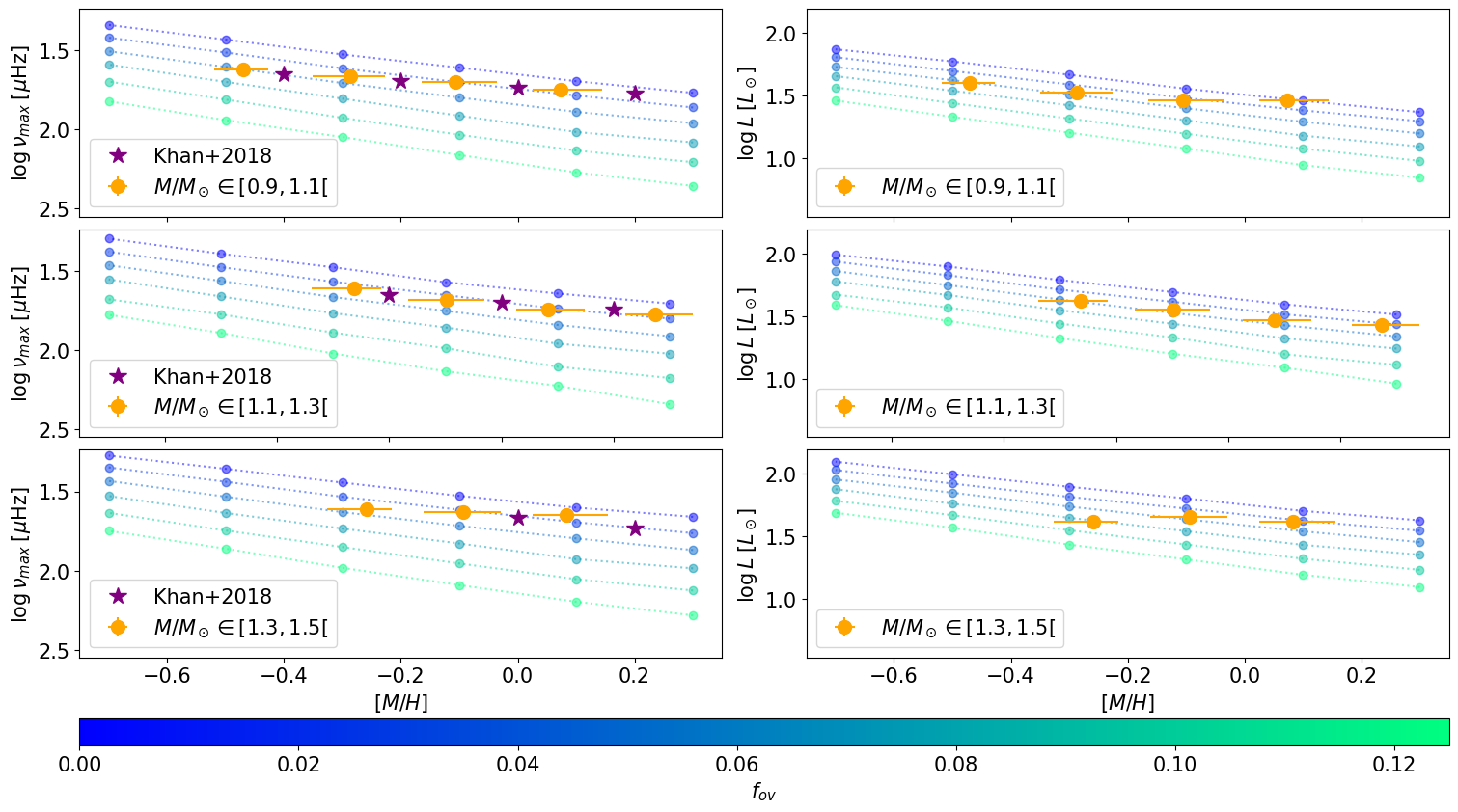}
    \caption{Location of the RGBb in $\log\numax$ (left) and $\log L$ (right), as a function of metallicity, for \textit{Kepler} stars divided in mass bins (orange points). The points coloured smoothly from green to blue represent the location of the RGBb in the same mass bins for different overshooting efficiencies. The colour-coding for $f_\mathrm{ov}$ is reported in the bottom horizontal bar. Our data are plotted as orange dots, while the results from \citealt{khan2018} are reported as purple stars (left panels).}
    \label{fig: Bump_trends_MH}
\end{figure*}

\begin{figure}
    \centering
    \includegraphics[width=\columnwidth]{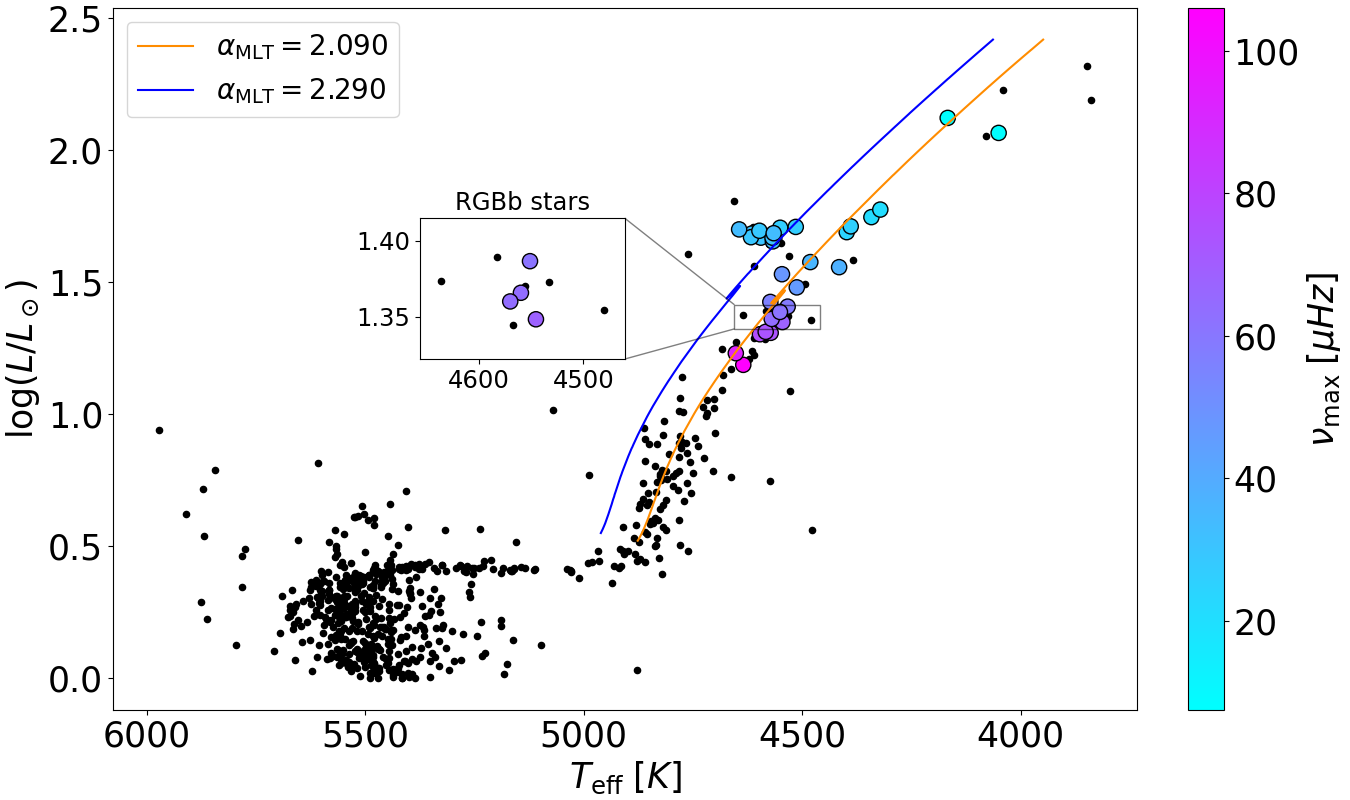}
    \caption{HRD of NGC 6791, with stars observed by \textit{Kepler} coloured by $\numax$. The inset zooms on the RGBb stars, including the four stars for which we have $\numax$ values from \textit{Kepler}. The orange and the blue lines correspond to the evolutionary tracks with $\aMLT = 2.090$ and $\aMLT = 2.290$, respectively, and no envelope overshooting.}
    \label{fig: NGC_6791}
\end{figure}

\begin{figure}
    \centering
    \includegraphics[width=\columnwidth]{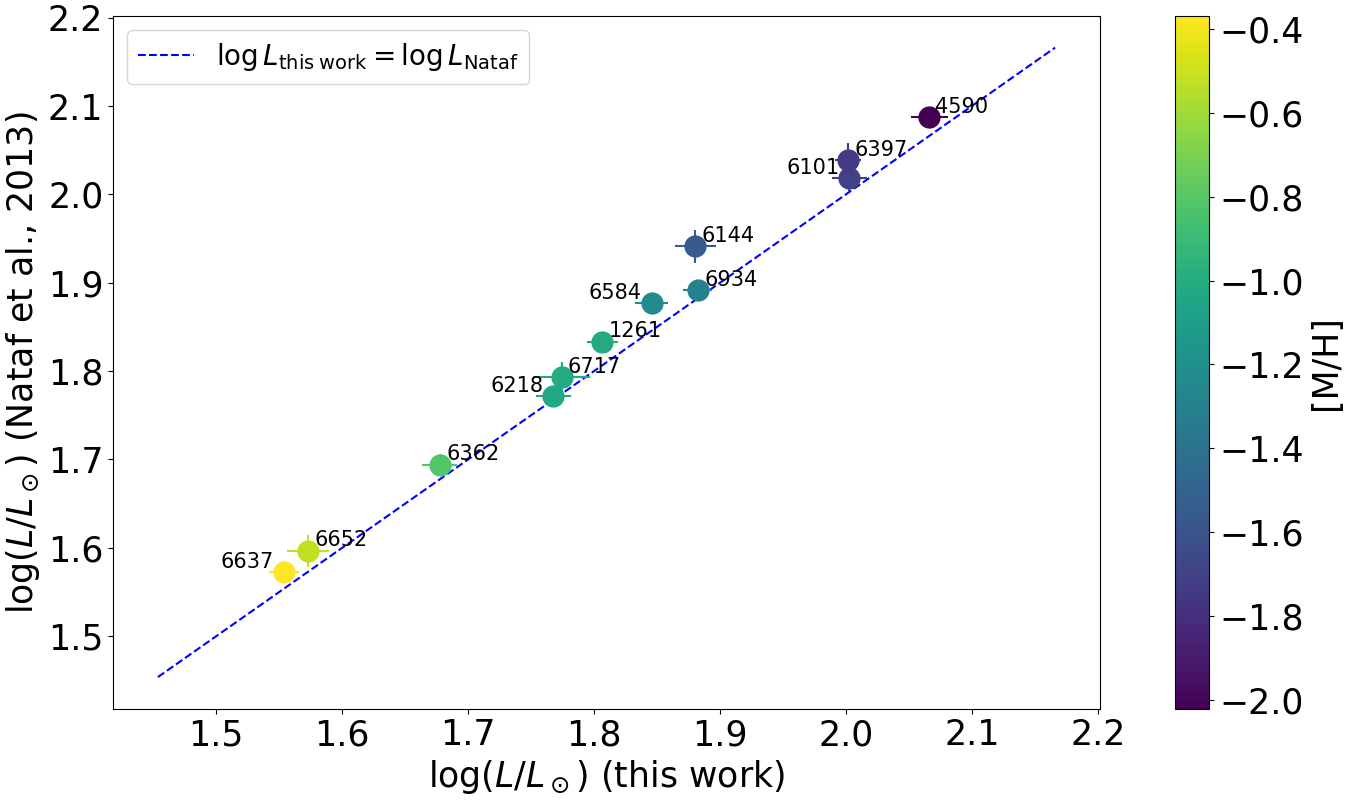}
    \caption{Comparison between our RGBb luminosities and the ones presented in \cite{nataf2013}.}
    \label{fig: Nataf_comparison}
\end{figure}

In this section we present our results on the determination of the RGBb locations and the overshooting efficiencies, adding a comparison between the bumps determined from luminosity and $\numax$, as well as a comment on the effects of changing $\aMLT$. Finally, we discuss a possible interpretation of the correlation between the overshooting efficiency and the metallicity.

\subsection{RGBb location}
\label{subsec: Results_RGBb_location}

We derived from the grid of models the variations of the RGBb location in luminosity and in $\numax$ in response to changes in stellar parameters (i.e. mass, metallicity, overshooting efficiency) and compared them to those emerging from the analysis of GCs and field stars. \par

Specifically, we observed from the evolutionary tracks that $\nu_\mathrm{max, RGBb}$ decreases with increasing stellar mass and decreasing metallicity, while the RGBb luminosity exhibits the opposite trends \citep[as already discussed by, e.g.,][]{fusipecci1990, cassisi&salaris1997, riello2003, dicecco2010, nataf2013, khan2018}. This phenomenon can be attributed to the fact that, in the range of masses explored,  more massive stars, as well as those with lower metallicity, are hotter and consequently have a shallower convective envelope and a more luminous RGBb. Consequently, since an increase in luminosity along the RGB corresponds to an increase in radius ($R$), and given that $\numax$ is strongly dependent on $R$ ($\numax \propto \mathrm{M}R^{-2}\Teff^{-1/2}$, see e.g. \citealt{brown1991, kjeldsen&bedding1995, belkacem2011}), we expect $\nu_\mathrm{max, RGBb}$ to exhibit trends opposite to those observed for luminosity. In addition, we found lower RGBb luminosities (or, equivalently, higher $\nu_\mathrm{max, RGBb}$ values) for higher values of the overshooting efficiency: this is due to the fact that a more efficient overshooting produces a deeper mixing zone, leading to a less luminous RGBb. \par

We inferred the RGBb luminosities for stellar clusters (see Tab.~\ref{tab:Clusters_properties} and Fig.~\ref{fig: logL_MH_GC}) and the RGBb location in $L$ and $\numax$ for field stars (see Tab.~\ref{tab: field_stars_bins}, Fig.~\ref{fig: Bump_trends_Mass} and Fig.~\ref{fig: Bump_trends_MH}), obtaining trends with mass and metallicity that broadly reflect the expectations from models. For the OC NGC 6791 we were also able to estimate the RGBb $\numax$, given that some giants belonging to that cluster, including four RGBb stars, have been observed by \textit{Kepler} and were included in our sample from the catalogue of \citealt{willett2025} (see Fig.~\ref{fig: NGC_6791}). Computing the $\numax$ median value of those four stars we obtained $\nu_\mathrm{max, RGBb} = 64^{+2}_{-1}\; \mu\mathrm{Hz}$. \par

Our results are also consistent with previous studies. In particular, we compared our RGBb positions in GCs with the ones presented in \citealt{nataf2013} (which we converted from V magnitude to $\log L$ with the bolometric corrections from \citealt{casagrande2014}). Despite a little systematic offset between the two works (the median difference between $\log L_\mathrm{Nataf}$ and $\log L_\mathrm{this\; work}$ is $0.020^{+0.013}_{-0.006}$), we found general agreement, as shown in Fig.~\ref{fig: Nataf_comparison}. We also found that the RGBb frequencies of our field stars are compatible with the ones obtained in the work by \citealt{khan2018} (see Fig.~\ref{fig: Bump_trends_MH}).
\subsection{Comparison between \texorpdfstring{$L_\mathrm{RGBb}$}{L\_RGBb} and \texorpdfstring{$\nu_\mathrm{max, RGBb}$}{nu\_max, RGBb}}
\label{subsec: L_numax_comparison}

\renewcommand{\arraystretch}{1.4}
\begin{table*}
\caption{Comparison between the three different measurements of $logL_\mathrm{RGBb}$ for field stars.}            
\label{tab:logL_comparison}     
\centering                          
\begin{tabular}{c c c c c}        
\hline \hline            
$\mathrm{M}\; [\Msun]$ & $\MH$ & $\log L_\mathrm{fit}\; [L_\odot]$ &  $\log L_{\numax}\; [L_\odot]$ & $\log L_\mathrm{scaling}\; [L_\odot]$\\
\hline                     
    $0.97^{+0.07}_{-0.04}$ & $-0.47^{+0.04}_{-0.05}$ & $1.60^{+0.02}_{-0.02}$ & $1.54^{+0.06}_{-0.05}$ & $1.56^{+0.04}_{-0.03}$ \\
    $1.01^{+0.06}_{-0.06}$ & $-0.29^{+0.06}_{-0.06}$ & $1.53^{+0.01}_{-0.01}$ & $1.52^{+0.06}_{-0.06}$ & $1.52^{+0.03}_{-0.03}$ \\
    $1.02^{+0.05}_{-0.05}$ & $-0.11^{+0.07}_{-0.06}$ & $1.46^{+0.01}_{-0.01}$ & $1.47^{+0.04}_{-0.04}$ & $1.46^{+0.03}_{-0.03}$ \\
    $1.04^{+0.04}_{-0.08}$ & $+0.07^{+0.07}_{-0.05}$ & $1.46^{+0.01}_{-0.01}$ & $1.42^{+0.05}_{-0.04}$ & $1.41^{+0.03}_{-0.04}$ \\
    $1.17^{+0.07}_{-0.05}$ & $-0.26^{+0.05}_{-0.07}$ & $1.62^{+0.01}_{-0.01}$ & $1.65^{+0.08}_{-0.04}$ & $1.65^{+0.03}_{-0.03}$ \\
    $1.20^{+0.06}_{-0.06}$ & $-0.10^{+0.07}_{-0.07}$ & $1.55^{+0.01}_{-0.01}$ & $1.55^{+0.06}_{-0.04}$ & $1.57^{+0.03}_{-0.03}$ \\
    $1.19^{+0.07}_{-0.06}$ & $+0.08^{+0.07}_{-0.06}$ & $1.47^{+0.01}_{-0.01}$ & $1.49^{+0.04}_{-0.05}$ & $1.48^{+0.03}_{-0.03}$ \\
    $1.18^{+0.06}_{-0.05}$ & $+0.27^{+0.07}_{-0.05}$ & $1.43^{+0.01}_{-0.01}$ & $1.43^{+0.03}_{-0.05}$ & $1.42^{+0.03}_{-0.03}$ \\
    $1.38^{+0.08}_{-0.06}$ & $-0.26^{+0.04}_{-0.07}$ & $1.62^{+0.02}_{-0.02}$ & $1.64^{+0.05}_{-0.04}$ & $1.72^{+0.03}_{-0.03}$ \\
    $1.39^{+0.06}_{-0.07}$ & $-0.09^{+0.06}_{-0.07}$ & $1.66^{+0.01}_{-0.01}$ & $1.66^{+0.05}_{-0.04}$ & $1.69^{+0.03}_{-0.03}$ \\
    $1.38^{+0.07}_{-0.06}$ & $+0.08^{+0.07}_{-0.06}$ & $1.62^{+0.01}_{-0.01}$ & $1.64^{+0.05}_{-0.03}$ & $1.65^{+0.03}_{-0.03}$ \\
\hline
\end{tabular}
\end{table*}
\renewcommand{\arraystretch}{1}

As mentioned above, we derived the RGBb location with two different observables: luminosity and frequency of maximum oscillation power. This allowed us to obtain two more measurements of $\log L_\mathrm{RGBb}$, to compare with the one coming from the fit ($\log L_\mathrm{fit}$). \par

The first additional measurement ($\log L_{\numax}$) comes from the median luminosity of the stars that were selected as RGBb stars according to their $\numax$ (as explained in section~\ref{sec: Methods}). In all the bins of field stars $\log L_\mathrm{fit}$ and $\log L_{\numax}$ are compatible, meaning that the fits in $\log L$ and $\log\numax$ classify as RGBb stars almost the same objects (see Table~\ref{tab:logL_comparison}). The second additional measurement ($\log L_\mathrm{scaling}$) has been derived from the asteroseismic scaling relations \citep[see e.g.][]{brown1991, frandsen2007, chaplin&miglio2013, miglio2016, miglio2021a} and, in particular, from the one that combines $M$, $L$, $\numax$ and $\Teff$:
\begin{equation}
    \frac{\mathrm{M}}{\Msun} = \biggl(\frac{L}{L_\odot}\biggr)\biggl(\frac{\numax}{\nu_{\mathrm{max}, \odot}}\biggr)\biggl(\frac{\Teff}{T_{\mathrm{eff}, \odot}}\biggr)^{-7/2}.
    \label{eq: astero_scaling_relation}
\end{equation}
Also in this case we found agreement with $\log L_\mathrm{fit}$, with the exception of the bin with $\mathrm{M}/\Msun \in [1.3,1.5[$ and $\MH \in [-0.4, -0.2[$ \si{dex}, for which the luminosity computed with eq.~\ref{eq: astero_scaling_relation} is higher (see Table~\ref{tab:logL_comparison}). \par

This bin presents, compared with the others, an anomalously low RGBb luminosity (see Figs.~\ref{fig: Bump_trends_Mass} and~\ref{fig: Bump_trends_MH}) and the evolution of its stars could be affected by effects such as core overshooting \citep{khan2018} and rotation \citep{charbonnel&lagarde2010, lagarde2012, beyer&white2024} during the MS, which start to arise in these intervals of mass and metallicity. We thus decided to exclude this bin from our analysis.

\subsection{Effects of changing the mixing length parameter}
\label{subsec: mixing_length}

\begin{figure}
    \centering
    \includegraphics[width=\columnwidth]{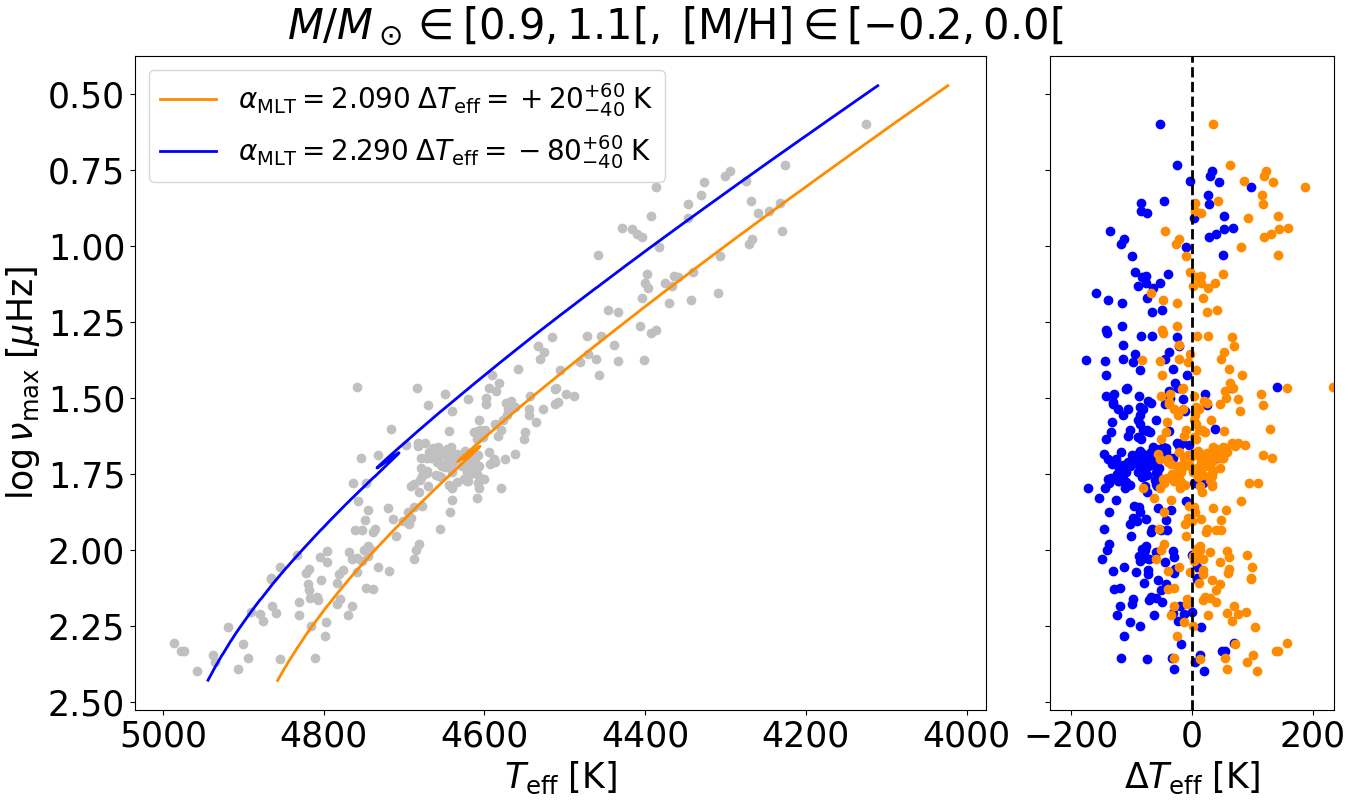}
    \caption{\textit{Left panel}: $\Teff-\log\numax$ diagram for field stars with $\mathrm{M}/\Msun \in [0.9, 1.1[$ and $\MH \in [-0.2, 0.0[$ \si{dex} (grey points) and corresponding evolutionary tracks with $\aMLT = 2.090$ (orange line) and $\aMLT = 2.290$ (blue line). \textit{Right panel}: Temperature differences between the data and the evolutionary tracks, with the same colour coding of the left panel.}
    \label{fig: Teff_diff}
\end{figure}

As mentioned in section~\ref{subsec: Data_Models}, we computed our grid of evolutionary tracks with two different values of $\aMLT$, which is the parameter that regulates the length of the mean free path and the size of the convective cells according to the Mixing Length Theory \citep{bohm-vitense1958}. The values we chose are $\aMLT = 2.290$, which comes from the solar calibration of our models, and $\aMLT = 2.090$, which gives a better fit to the temperatures from APOGEE, as explained below. \par

Variations of this parameter affect stellar radii and effective temperatures. Specifically, an increment of $\aMLT$ produces stellar models with higher temperatures and smaller radii \citep{cox&giuli1968}. This implies variations in $\numax$ and, to a lesser extent,  in luminosities (since the increment in effective temperature and the decrement in radius tend to compensate each other). Since a modification of $\aMLT$ produces different variations of the RGBb location in $L$ and $\numax$, we expect to find also different overshooting efficiencies for different values of $\aMLT$, with different impacts on $\fov^{L}$ and $\fov^{\numax}$. In this work we found the best agreement between $\fov^{L}$ and $\fov^{\numax}$ for $\aMLT =2.090$ (see section~\ref{subsec: Overshooting_calibration}). \par

Crucially, we found that this value of $\aMLT$ also significantly improves the agreement between the predicted and observed \Teff. Specifically, we computed the median $\Delta\Teff = T_\mathrm{eff,data}-T_\mathrm{eff,track}$ for each bin of field stars (an example is reported in Fig.~\ref{fig: Teff_diff}), finding that the tracks with $\aMLT = 2.090$ adapt better to the data in all our bins. In addition, we computed the median of all those quantities, finding $\Delta\Teff = +30^{+04}_{-20}\; \mathrm{K}$ for $\aMLT=2.090$ and $\Delta\Teff = -60^{+05}_{-20}\; \mathrm{K}$ for $\aMLT=2.290$. A similar conclusion is reached for the OC NGC 6791, as shown in Fig.~\ref{fig: NGC_6791}

\subsection{Envelope overshooting calibration}
\label{subsec: Overshooting_calibration}
 
   \begin{figure*}
   \centering
   \includegraphics[width=8.5cm]{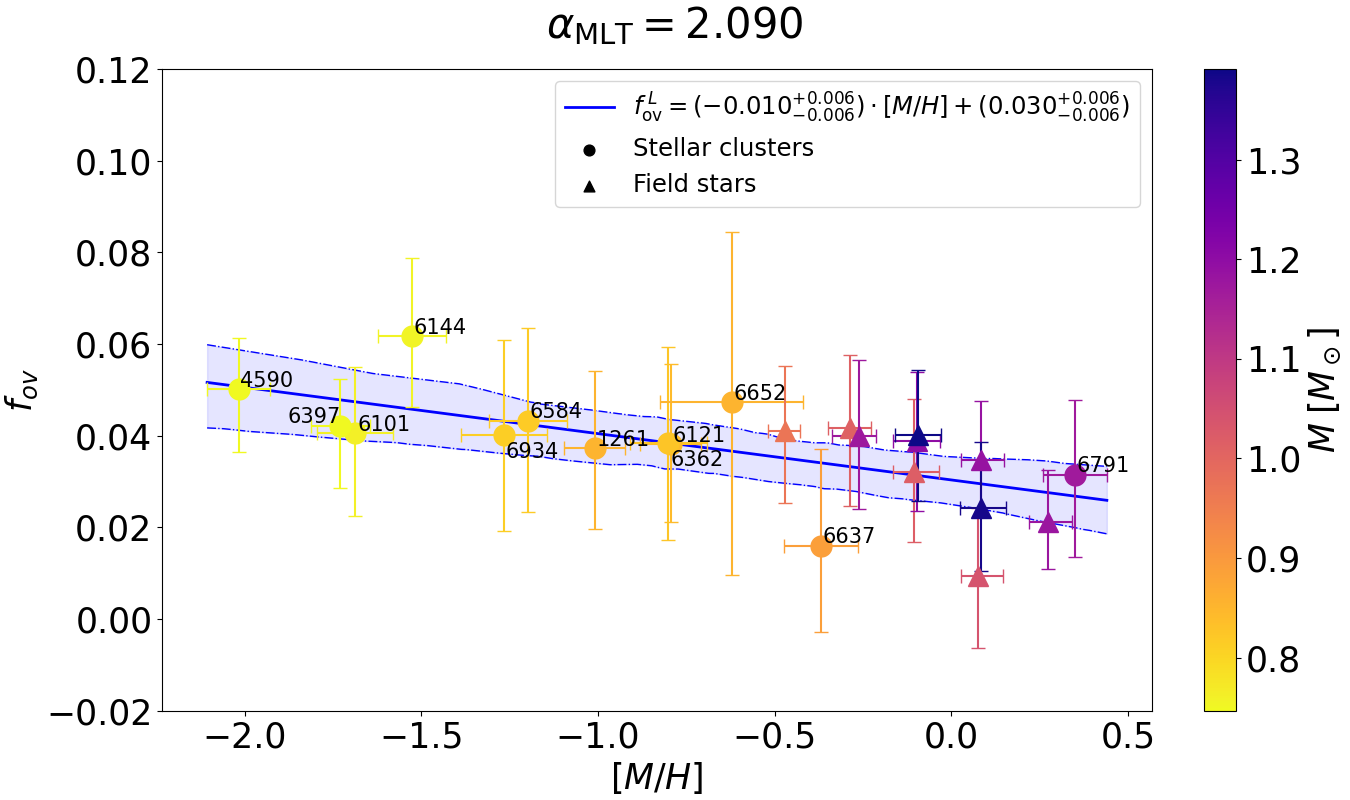} 
   \includegraphics[width=8.5cm]{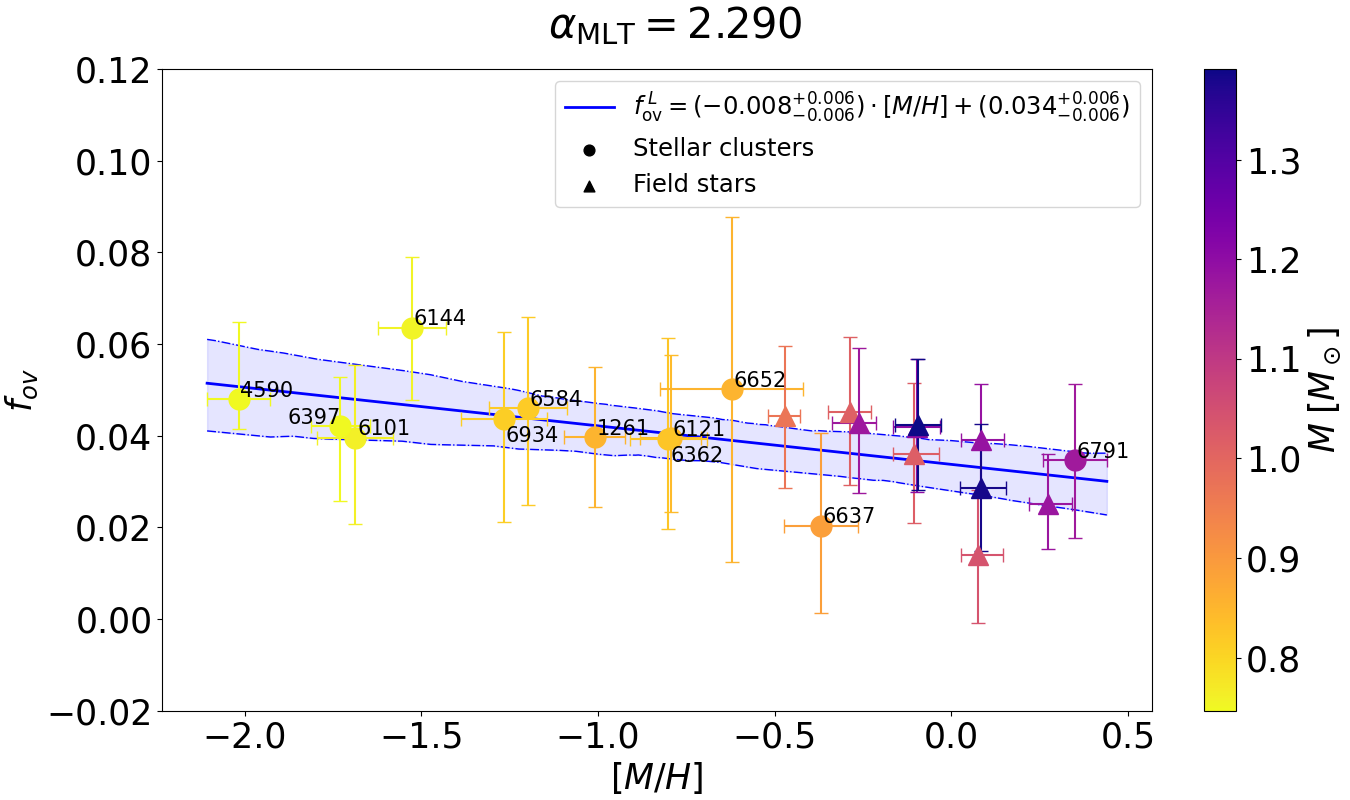} 
   \caption{Overshooting efficiency derived from $L_\mathrm{RGBb}$ as a function of $\MH$, both for $\aMLT = 2.090$ (left panel) and $\aMLT = 2.290$ (right panel). Clusters are represented as circles, while field stars as triangles. The blue lines correspond to the linear fits. Points are coloured by stellar mass.}
    \label{fig: fov_MH_L}
    \end{figure*} 
  
   \begin{figure*}
   \centering
   \includegraphics[width=8.5cm]{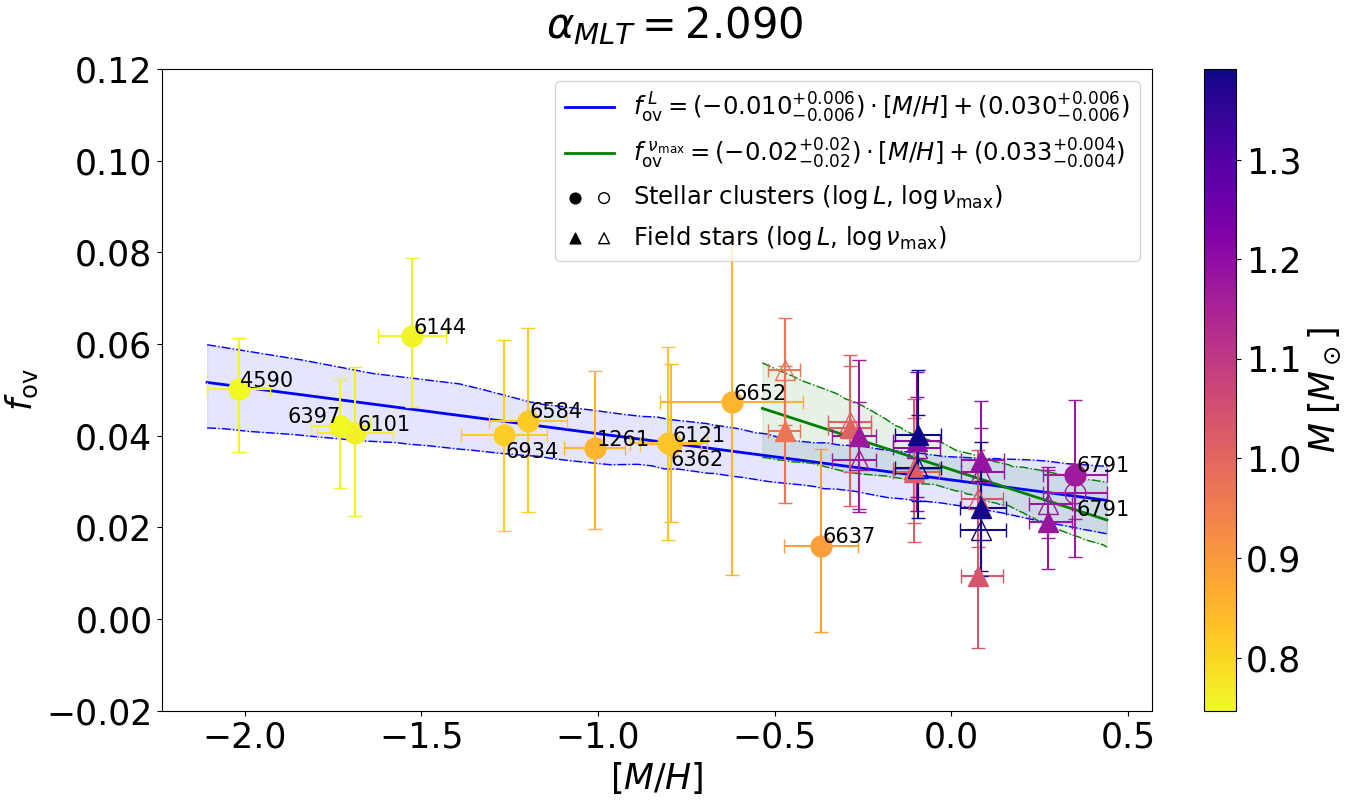} 
   \includegraphics[width=8.5cm]{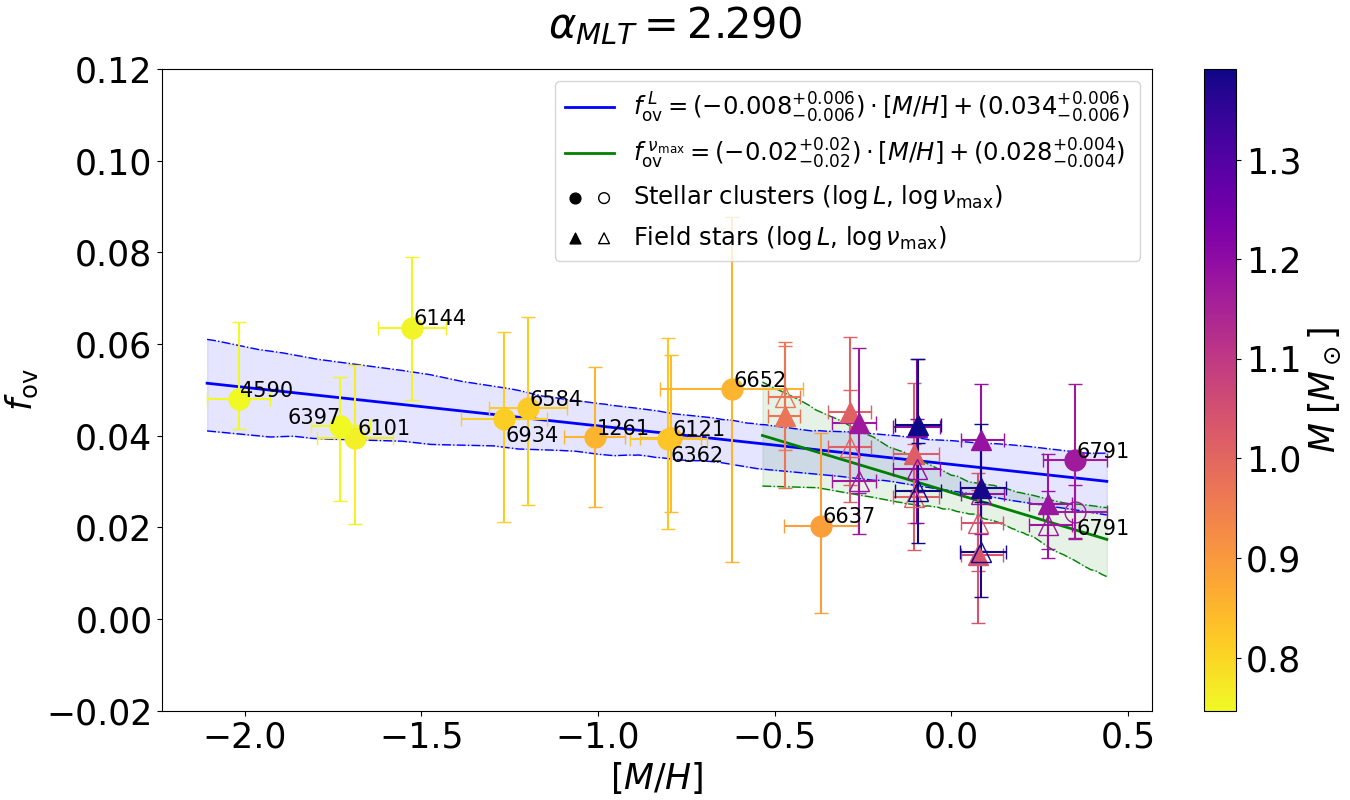} 
   \caption{Same as Fig.~\ref{fig: fov_MH_L}, but with the addition of the overshooting efficiencies derived from $\nu_\mathrm{max, RGBb}$ (open symbols) and the relative fits (green lines).}
    \label{fig: fov_MH}
    \end{figure*} 
   
   \begin{figure*}
   \centering
   \includegraphics[width=8.5cm]{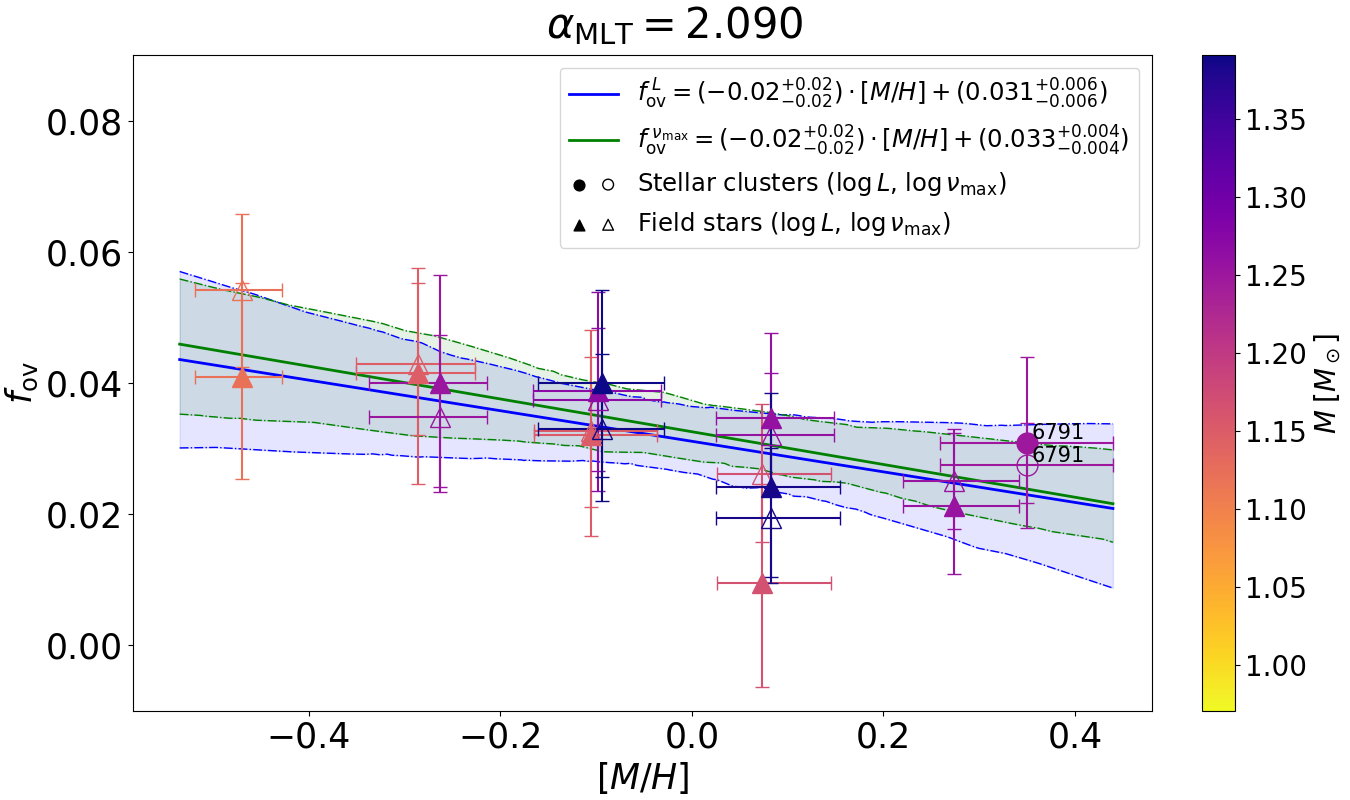} 
   \includegraphics[width=8.5cm]{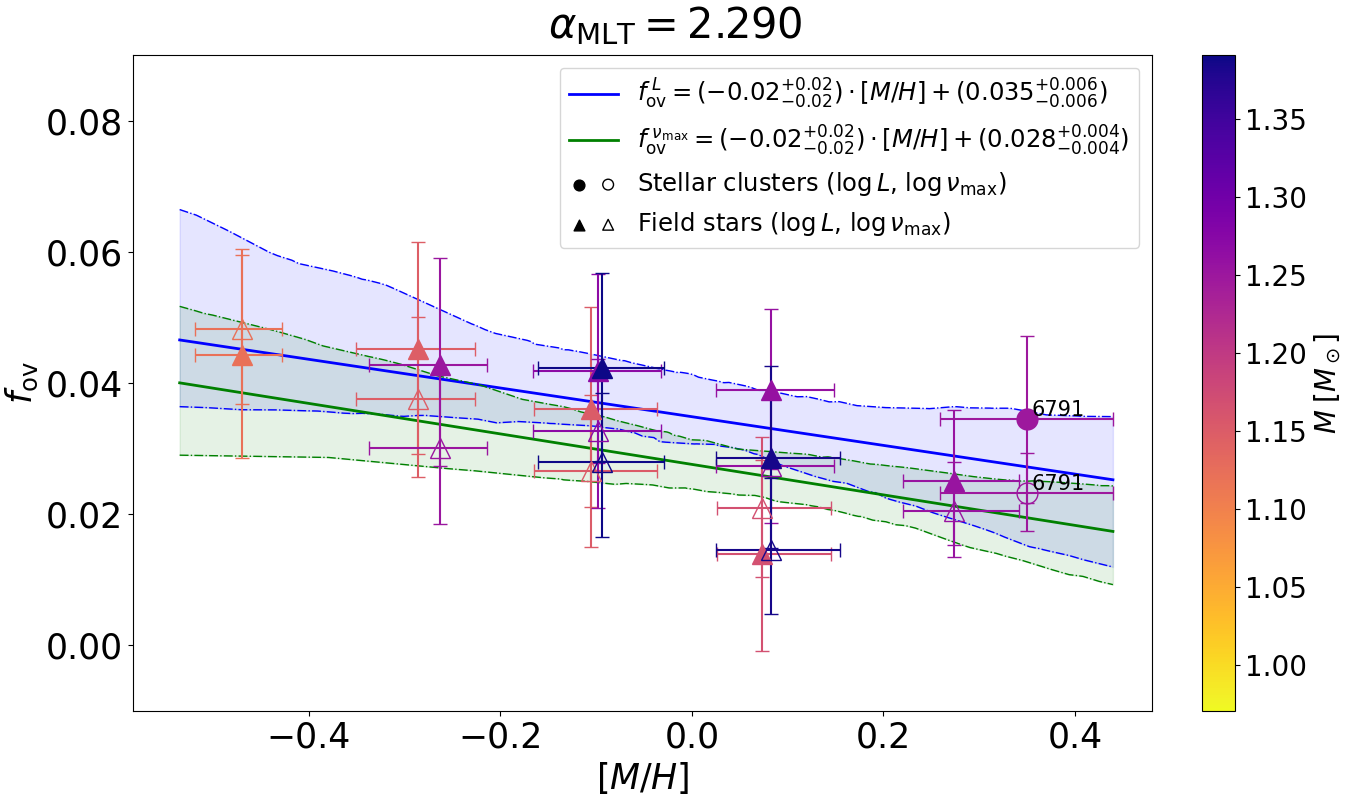} 
   \caption{Same as Fig.~\ref{fig: fov_MH}, but only with points for which we computed both $\fov^{L}$ and $\fov^{\numax}$.}
    \label{fig: fov_MH_no_globulars}
    \end{figure*} 

\begin{figure}
    \centering
    \includegraphics[width=\columnwidth]{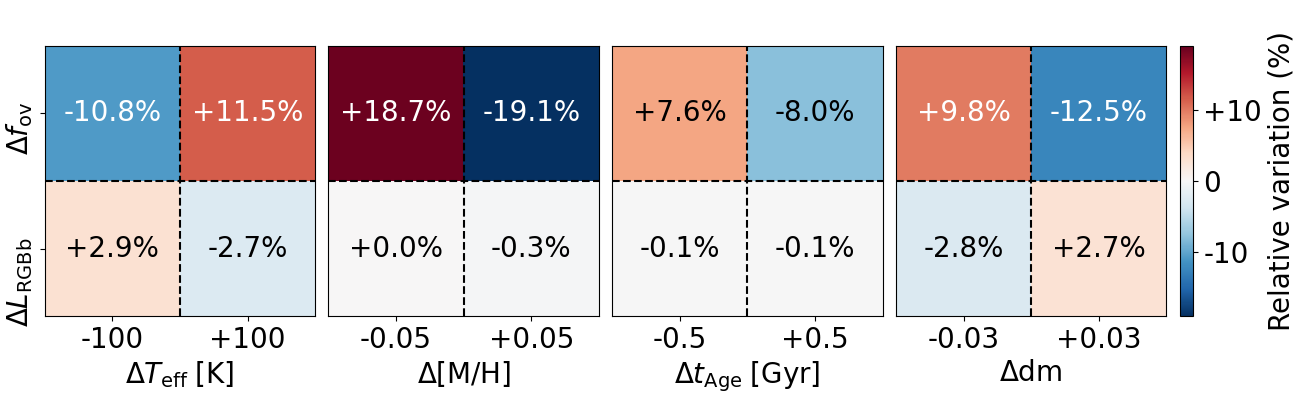}
    \caption{Relative variations of RGBb luminosity and overshooting efficiency for the GC NGC 6362 produced by variations of effective temperature ($\Teff$), metallicity ($\MH$), age ($t_\mathrm{Age}$) and distance modulus (dm).}
    \label{fig: Heatmap}
\end{figure}

\renewcommand{\arraystretch}{1.4}
\begin{table*}
\caption{Interpolation results}         
\label{tab:interpolation_results}     
\centering                          
\begin{tabular}{c c c | c c | c c}        
\hline \hline           
\multirow{2}{*}{NGC} & \multirow{2}{*}{$\mathrm{M}\; [\Msun]$} & \multirow{2}{*}{$\MH$} & \multicolumn{2}{c|}{$\aMLT = 2.090$} & \multicolumn{2}{c}{$\aMLT = 2.290$} \\
& & & $\fov^{L}$ & $\fov^{\numax}$ & $\fov^{L}$ & $\fov^{\numax}$ \\
\hline    
\multicolumn{7}{c}{Clusters} \\
\hline
    1261 & $0.85^{+0.01}_{-0.01}$ & $-1.01^{+0.09}_{-0.09}$ & $0.037^{+0.017}_{-0.018}$ & --- & $0.040^{+0.015}_{-0.015}$ & --- \\
    4590 & $0.75^{+0.01}_{-0.01}$ & $-2.02^{+0.09}_{-0.09}$ & $0.050^{+0.011}_{-0.014}$ & --- & $0.048^{+0.017}_{-0.007}$ & --- \\
    6101 & $0.76^{+0.01}_{-0.01}$ & $-1.69^{+0.11}_{-0.11}$ & $0.041^{+0.014}_{-0.018}$ & --- & $0.040^{+0.016}_{-0.019}$ & --- \\
    6121 & $0.84^{+0.01}_{-0.01}$ & $-0.80^{+0.09}_{-0.09}$ & $0.038^{+0.017}_{-0.017}$ & --- & $0.039^{+0.018}_{-0.016}$ & --- \\
    6144 & $0.76^{+0.01}_{-0.01}$ & $-1.53^{+0.10}_{-0.10}$ & $0.062^{+0.017}_{-0.015}$ & --- & $0.064^{+0.015}_{-0.016}$ & --- \\
    6362 & $0.83^{+0.01}_{-0.01}$ & $-0.80^{+0.11}_{-0.11}$ & $0.038^{+0.021}_{-0.021}$ & --- & $0.039^{+0.022}_{-0.020}$ & --- \\
    6397 & $0.75^{+0.01}_{-0.01}$ & $-1.73^{+0.08}_{-0.08}$ & $0.042^{+0.010}_{-0.014}$ & --- & $0.042^{+0.011}_{-0.016}$ & --- \\
    6584 & $0.82^{+0.01}_{-0.01}$ & $-1.20^{+0.11}_{-0.11}$ & $0.043^{+0.020}_{-0.020}$ & --- & $0.046^{+0.020}_{-0.021}$ & --- \\
    6637 & $0.89^{+0.02}_{-0.02}$ & $-0.37^{+0.10}_{-0.10}$ & $0.016^{+0.021}_{-0.019}$ & --- & $0.020^{+0.020}_{-0.019}$ & --- \\
    6652 & $0.85^{+0.02}_{-0.02}$ & $-0.62^{+0.20}_{-0.20}$ & $0.047^{+0.037}_{-0.038}$ & --- & $0.050^{+0.038}_{-0.038}$ & --- \\
    6934 & $0.81^{+0.01}_{-0.01}$ & $-1.27^{+0.12}_{-0.12}$ & $0.040^{+0.021}_{-0.021}$ & --- & $0.044^{+0.019}_{-0.022}$ & --- \\
    6791 & $1.17^{+0.02}_{-0.02}$ & $+0.35^{+0.09}_{-0.09}$ & $0.031^{+0.016}_{-0.018}$ & $0.027^{+0.006}_{-0.006}$ & $0.035^{+0.017}_{-0.017}$ & $0.023^{+0.006}_{-0.006}$ \\
\hline    
\multicolumn{7}{c}{Field stars} \\
\hline
    --- & $0.97^{+0.07}_{-0.04}$ & $-0.47^{+0.04}_{-0.05}$ & $0.041^{+0.014}_{-0.016}$ & $0.054^{+0.012}_{-0.012}$ & $0.044^{+0.015}_{-0.016}$ & $0.048^{+0.012}_{-0.011}$ \\
    --- & $1.01^{+0.06}_{-0.06}$ & $-0.29^{+0.06}_{-0.06}$ & $0.042^{+0.016}_{-0.017}$ & $0.043^{+0.012}_{-0.011}$ & $0.045^{+0.016}_{-0.016}$ & $0.038^{+0.012}_{-0.012}$ \\
    --- & $1.02^{+0.05}_{-0.05}$ & $-0.11^{+0.07}_{-0.06}$ & $0.032^{+0.016}_{-0.015}$ & $0.033^{+0.011}_{-0.012}$ & $0.036^{+0.015}_{-0.015}$ & $0.027^{+0.012}_{-0.011}$ \\
    --- & $1.04^{+0.04}_{-0.08}$ & $+0.07^{+0.07}_{-0.05}$ & $0.009^{+0.015}_{-0.016}$ & $0.026^{+0.011}_{-0.010}$ & $0.014^{+0.014}_{-0.015}$ & $0.021^{+0.011}_{-0.011}$ \\
    --- & $1.17^{+0.07}_{-0.05}$ & $-0.26^{+0.05}_{-0.07}$ & $0.040^{+0.017}_{-0.016}$ & $0.035^{+0.013}_{-0.011}$ & $0.043^{+0.016}_{-0.015}$ & $0.030^{+0.012}_{-0.012}$ \\
    --- & $1.20^{+0.06}_{-0.06}$ & $-0.10^{+0.07}_{-0.07}$ & $0.039^{+0.015}_{-0.015}$ & $0.037^{+0.011}_{-0.011}$ & $0.042^{+0.015}_{-0.014}$ & $0.033^{+0.011}_{-0.012}$ \\
    --- & $1.19^{+0.07}_{-0.06}$ & $+0.08^{+0.07}_{-0.06}$ & $0.035^{+0.013}_{-0.011}$ & $0.032^{+0.009}_{-0.008}$ & $0.039^{+0.012}_{-0.012}$ & $0.027^{+0.010}_{-0.009}$ \\
    --- & $1.18^{+0.06}_{-0.05}$ & $+0.27^{+0.07}_{-0.05}$ & $0.021^{+0.011}_{-0.010}$ & $0.025^{+0.008}_{-0.007}$ & $0.025^{+0.011}_{-0.010}$ & $0.020^{+0.008}_{-0.007}$ \\
    --- & $1.39^{+0.06}_{-0.07}$ & $-0.09^{+0.06}_{-0.07}$ & $0.040^{+0.014}_{-0.014}$ & $0.033^{+0.011}_{-0.011}$ & $0.042^{+0.015}_{-0.014}$ & $0.028^{+0.011}_{-0.011}$ \\
    --- & $1.38^{+0.07}_{-0.06}$ & $+0.08^{+0.07}_{-0.06}$ & $0.024^{+0.014}_{-0.014}$ & $0.019^{+0.011}_{-0.010}$ & $0.029^{+0.014}_{-0.014}$ & $0.014^{+0.011}_{-0.010}$ \\
\hline
\end{tabular}
\tablefoot{Columns are: NGC number (only for clusters), mass (derived with eq.~\ref{eq: mass-age-mh_relation} for clusters, median mass in the bin for field stars), [M/H], and overshooting efficiencies derived from 
$\log L_\mathrm{RGBb}$ and $\log\nu_\mathrm{max, RGBb}$ (the latter only for field stars and the OC NGC 6791), both for $\aMLT = 2.090$ and $\aMLT = 2.290$.}
\end{table*}
\renewcommand{\arraystretch}{1}

\renewcommand{\arraystretch}{1.4}
\begin{table*}
\caption{Results of the fits on the $\fov$-$\MH$ plane, each one with the corresponding $\aMLT$ value and metallicity range.}            
\label{tab:fov_MH_fit_coefficients}     
\centering                          
\begin{tabular}{lcccc}        
\hline \hline        
  & $\aMLT$ & $\MH$ range & $m$ [\si{dex}$^{-1}$] & $q$ \\
\hline
    \multirow{2}{*}{$\fov^{L}$} & 2.090 & [-2.02, +0.35] &  $-0.010^{+0.006}_{-0.006}$& $+0.030^{+0.006}_{-0.006}$ \\
    & 2.290 & [-2.02, +0.35] & $-0.008^{+0.006}_{-0.006}$& $+0.034^{+0.006}_{-0.006}$ \\
\hline
    \multirow{2}{*}{$\fov^{\numax}$} & 2.090 & [-0.5, +0.35] & $-0.02^{+0.02}_{-0.02}$& $+0.033^{+0.004}_{-0.004}$ \\
    & 2.290 & [-0.5, +0.35] & $-0.02^{+0.02}_{-0.02}$& $+0.028^{+0.004}_{-0.004}$ \\
\hline
    \multirow{2}{*}{$\fov^{L}$} & 2.090 & [-0.5, +0.35] & $-0.02^{+0.02}_{-0.02}$& $+0.031^{+0.006}_{-0.006}$ \\
    & 2.290 & [-0.5, +0.35] & $-0.02^{+0.02}_{-0.02}$& $+0.035^{+0.006}_{-0.006}$ \\
\hline                       
\end{tabular}
\end{table*}
\renewcommand{\arraystretch}{1}

As already mentioned in section~\ref{sec: Methods}, after deriving the RGBb luminosities and $\numax$, we performed an interpolation to infer the overshooting efficiencies for both stellar clusters and field stars, in order to look for a correlation with $\MH$. \par

We obtained overshooting efficiencies $\fov$ spanning from $0.009^{+0.015}_{-0.016}$ to $0.064^{^+0.015}_{-0.016}$ and, in particular, not greater than $0.054^{+0.012}_{-0.012}$ for field stars (see Table~\ref{tab:interpolation_results}). This result is consistent with the constraints presented in \cite{khan2018}, where an upper limit of $\sim 0.05$ for $\fov$ in field stars with $\mathrm{M}/\Msun \in [0.9, 1.5]$ and $\MH \in [-0.5, +0.3]$ \si{dex} is found. \par

Regarding the correlation between $\fov$ and $\MH$, we first computed the Spearman Correlation Index ($\rho_\mathrm{S}$, see \citealt{spearman1904}), which measures the rank of correlation between two variables. We found $\rho_{\mathrm{S}, 2.090} = -0.718$ and $\rho_{\mathrm{S}, 2.290} = -0.607$, both with p-values lower than $1\%$, indicating an anti-correlation between $\fov$ and $\MH$. We consequently tried to fit the points with a linear relation of the kind $\fov = m\cdot\MH+q$, finding a negative slope in all cases we analysed (see Tab.~\ref{tab:fov_MH_fit_coefficients}). \par

This process was repeated for both values of $\aMLT$ (2.090, 2.290). Although we consider the results obtained from the interpolation along the grid with $\aMLT = 2.090$ to be more reliable, as we fitted the effective temperatures (see section~\ref{subsec: mixing_length}), we also present the results obtained with $\aMLT = 2.290$. The behaviour of the relation is similar in both cases, indicating that our conclusions are robust and not significantly affected by this systematic effect. \par

Starting from the overshooting efficiency derived from the bump luminosity ($\fov^{L}$), which has been computed for $\MH$ between -2.02 \si{dex} and +0.35 \si{dex}, we found a relation with a weakly negative slope, steeper for $\aMLT = 2.090$ (see Fig.~\ref{fig: fov_MH_L} and Tab.~\ref{tab:fov_MH_fit_coefficients}). We performed a t-Student test \citep{student1908, fornasini2008} on our fit results, using the Python package \verb|statsmodels| \citep{seabold&perktold2010}. We found in both cases that the \textit{p-value} is lower than 1\%, meaning that the slope is significant and that we can reject the hypothesis of a flat relation. \par

For the overshooting efficiency derived from $\nu_\mathrm{max, RGBb}$ ($\fov^{\numax}$, obtained for field stars and the OC NGC 6791, for $\MH$ between -0.5 \si{dex} and +0.35 \si{dex}), we found the same slope for the two different values of $\aMLT$, steeper than the one related to $\fov^{L}$ (see Fig.~\ref{fig: fov_MH} and Tab.~\ref{tab:fov_MH_fit_coefficients}).
In both cases, $\aMLT = 2.090$ and $\aMLT = 2.290$, $\fov^{L}$ and $\fov^{\numax}$ are compatible within $1\sigma$ (see Table~\ref{tab:interpolation_results}) and, if we consider only the metallicity range where we have measured both the two different $\fov$ ($\MH \in [-0.5, +0.35]$ \si{dex}), we find that $\fov^{L}$ and $\fov^{\numax}$ are related to $\MH$ with the same slope, meaning that extending the metallicity range with the inclusion of GCs implies a flattening of the relation. In addition, we observe that when the interpolation is performed along the grid with $\aMLT = 2.090$, the shift in $\fov$ between the two relations is smaller (see Fig.~\ref{fig: fov_MH_no_globulars} and Tab.~\ref{tab:fov_MH_fit_coefficients}). \par

The flattening of the relation at low metallicities made us consider the possibility of fitting our data with a broken-line function. 
To see whether this model adapts to our data better than the single straight line, we performed again the t-Student test using the Python library \verb|piecewise-regression| \citep{pilgrim2021} and we found that the hypothesis of no breakpoints cannot be rejected and that the relation with no breakpoints (i.e. the straight line) is the preferable one. \par

To investigate the primary contributors to the current uncertainty in the inferred overshooting efficiency, we used the globular cluster NGC 6362 as a case study. This cluster has intermediate metallicity ($\MH = -0.80 \pm 0.11$), which lies within the $\MH$ range of our sample. We applied to several parameters variations of the order of their typical observational uncertainties: $\Teff$ by $\pm 100; \si{K}$, $\MH$ by $\pm 0.05\; \si{dex}$, age by $\pm 0.5\; \si{Gyr}$, and distance modulus (dm) by $\pm 0.03$. For each variation, we computed the effects on the inferred $\log g$, the bolometric corrections (BCs), and, consequently, the RGBb luminosity of the cluster. We then interpolated along the model grid using the new values of age ($t_\mathrm{Age}$), metallicity ($[\ch{M}/\ch{H}]$), and $\log L_\mathrm{RGBb}$ to compute the associated variations in $\fov$. Our results show that metallicity is the dominant contributor to the variation in $\fov$, accounting for approximately 19\%, followed by $\Teff$ and distance modulus (each contributing about 11\%), and age ($t_\mathrm{Age}$), which contributes around 8\%. On the other hand, $\Teff$ and the distance modulus have a greater impact on the RGBb luminosity than $\MH$ and $t_\mathrm{Age}$ (see Fig.~\ref{fig: Heatmap}). These findings indicate that to improve the precision of the overshooting efficiency measurements, a more accurate determination of metallicity is crucial, along with better constraints on temperature and distance. 

\subsection{A possible interpretation}
\label{subsec: Entrainment_law}

\begin{figure}
    \centering
    \includegraphics[width=\columnwidth]{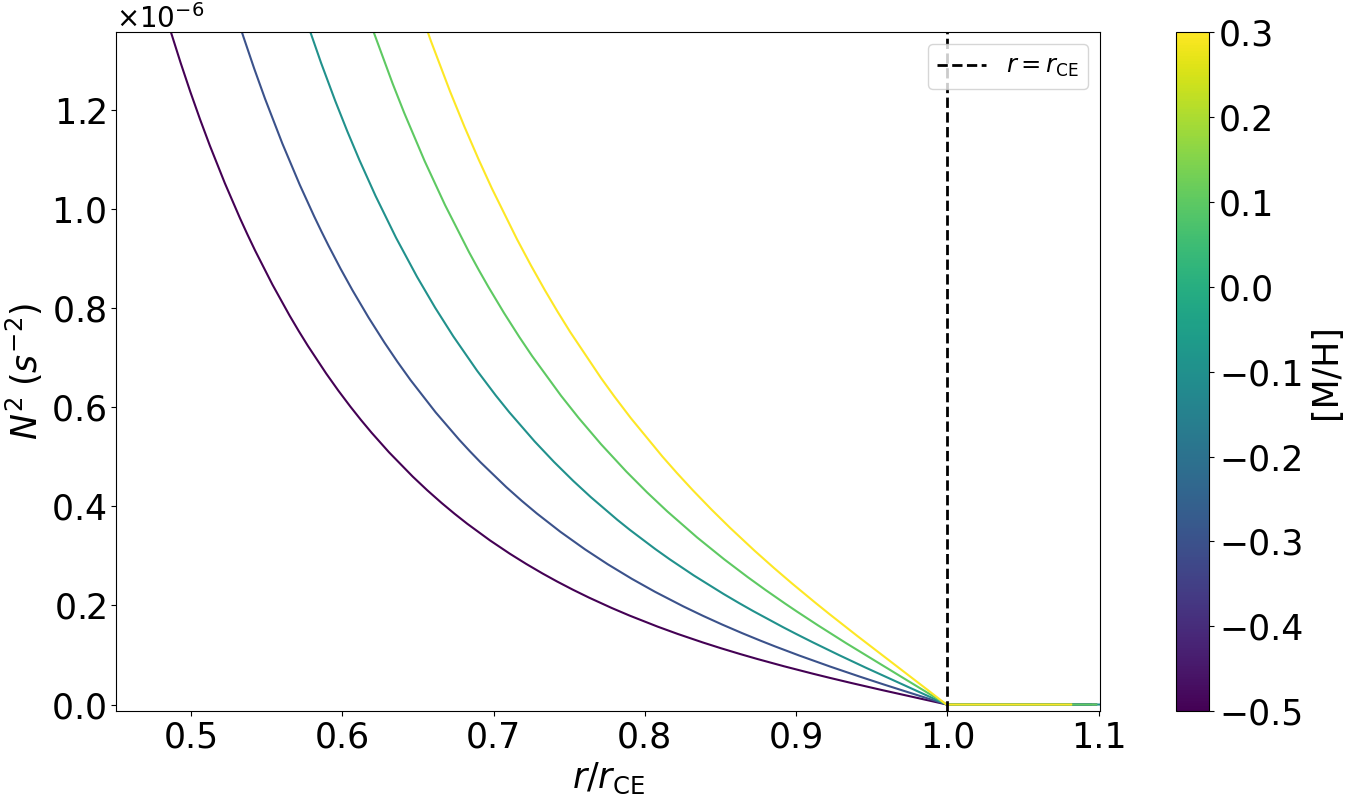}
    \caption{Profiles of the squared Brunt-Väisälä frequency $N^2$ for models with $\mathrm{M} = 1 \Msun$ and $\MH = -0.5, -0.3, -0.1, +0.1, +0.3$ \si{dex}. The \textit{x-axis} is normalised by the position $r_\mathrm{CE}$ of the convective border, indicated by the vertical black dashed line}
    \label{fig: N2_profiles}
\end{figure}

Given the anti-correlation between overshooting efficiency and metallicity (see Sec.~\ref{subsec: Overshooting_calibration}), 
we investigated, as done in \citealt{khan2021}, possible correlations of $\MH$ with the Brunt-Väisälä frequency, the characteristic frequency of internal gravity waves (i.e. oscillations driven by buoyancy), defined as:
\begin{equation}
    N^2 = g\biggl(\frac{1}{\Gamma_1}\frac{\mathrm{d}\ln P}{\mathrm{d}r}-\frac{\mathrm{d}\ln\rho}{\mathrm{d}r}\biggr),
    \label{eq: Brunt_Vaisala_def}
\end{equation}
where $g$ is the gravitational acceleration, $\Gamma_1 = (\partial\ln P/\partial\ln\rho)_\mathrm{ad}$ is one of the three adiabatic exponents, and $P$ and $\rho$ are pressure and density, respectively. Assuming the ideal gas law for a fully ionised gas, the square of the Brunt-Väisälä frequency can also be expressed in terms of the adiabatic ($\nabla_\mathrm{ad} = (\partial\ln T/\partial\ln P)_\mathrm{ad}$) and radiative ($\nabla_\mathrm{rad} = \mathrm{d}\ln T/\mathrm{d}\ln P$) temperature gradients, and the mean molecular weight gradient ($\nabla_\mu = \mathrm{d}\ln\mu/\mathrm{d}\ln P$), with the following approximation:
\begin{equation}
    N^2 \sim \frac{g^2\rho}{P}(\nabla_\mathrm{ad}-\nabla_\mathrm{rad}+\nabla_\mu),
    \label{eq: Brunt_Vaisala_approx}
\end{equation}
which shows that $N^2$ is negative in convective zones and positive in radiative zones. From our models, we observe that the profile of the squared Brunt-Väisälä frequency near the convective border is steeper at higher metallicities (see Fig.~\ref{fig: N2_profiles}). \par

This has implications on the bulk Richardson number, $Ri_\mathrm{B}$ \citep{lettau&davidson1957}, a quantification of the stiffness of the convective boundary: low values ($\sim 10$) of $Ri_\mathrm{B}$ allows convective boundary mixing, while high values ($\sim 10^4$) are associated to a stiff boundary that inhibits convective boundary mixing \citep{cristini2016}. Mathematically, $Ri_\mathrm{B}$ corresponds to the ratio between the potential energy of restoration of the convective boundary and the kinetic energy of turbulent eddies:
\begin{equation}
    Ri_\mathrm{B} = \frac{\Delta B \mathcal{L}}{v^{2}_{c}/2},
    \label{eq: Richardson_number}
\end{equation}
where $\mathcal{L}$ is a length scale associated to the turbulent motion while $\Delta B$ is the buoyancy jump over a distance $\Delta r$ from the boundary position $r_\mathrm{B}$:
\begin{equation}
    \Delta B = \int_{r_\mathrm{B}-\Delta r}^{r_\mathrm{B}+\Delta r} N^2 dr.
    \label{eq: buoyancy_jump}
\end{equation}

From the last two equations, we can deduce that steeper profiles of $N^2$ produce higher buoyancy jumps and, consequently, higher values of $Ri_\mathrm{B}$, hence, steeper convective boundaries. An analogous connection between $Ri_\mathrm{B}$ (hence, the profile of $N^2$) and the extension of the mixing zone is found in \citealt{meakin&arnett2007}, which states that the entrainment coefficient $E$, i.e. the ratio between the time rate of change of the boundary position through turbulent entrainment \citep{kantha1977, strang&fernando2001}, $v_e$, and the rms turbulent velocity of the fluid elements, $v_c$, is inversely proportional to a power of the bulk Richardson number:
\begin{equation}
    E = \frac{v_e}{v_c} = ARi_\mathrm{B}^{-n}.
    \label{eq: entrainment_law}
\end{equation}
Typical values of $A$ and $n$ are in the ranges $[0.1;0.5]$ and $[1.00;1.75]$, respectively \citep{meakin&arnett2007}. \par

We can thus suppose that the decrease of the overshooting efficiency with increasing metallicity could be associated to the steepening of the squared Brunt-Väisälä profile below the convective zone.

\section{Summary and conclusions}
\label{sec: Conclusions}

In this paper, we extended the work described in \cite{khan2018}, with the aim of calibrating the efficiency of convective-envelope overshooting in RGB stars. \par

First, we derived the RGBb location in $\log\numax$ and $\log L$ for the widest mass and metallicity domain explored until now, including both stellar clusters and field stars. We found that the RGBb luminosity decreases with increasing metallicity and decreasing mass and, contextually, the RGBb $\numax$ follows the opposite trends (see Sec.~\ref{subsec: Results_RGBb_location}), confirming results from previous works \citep{nataf2013, khan2018}. \par

After that, from the comparison between the RGBb locations in our datasets and in our grid of evolutionary tracks, we derived overshooting efficiencies ($\fov$, i.e. the rate, in units of $H_\mathrm{P}$, at which the diffusion coefficient decays exponentially getting further from the convective envelope, see Sec.~\ref{subsec: Data_Models}) for stellar clusters and field stars and studied a correlation of this quantity with metallicity, measuring overshooting efficiencies ranging from $0.009^{+0.015}_{-0.016}$ to $0.062^{^+0.017}_{-0.015}$. We found that $\fov$ decreases with increasing metallicity and that the most suitable model to describe this trend in our data is a linear relation with slope $(-0.010\pm0.006)\; \si{dex}^{-1}$ (see Sec.~\ref{subsec: Overshooting_calibration}), which is significative according to the t-Student test \citep{student1908, fornasini2008}. 
We then explored the behaviour of the Brunt-Väisälä frequency under variations of $[\ch{M}/\ch{H}]$, finding that at higher metallicities $N^2$ presents steeper profiles, increasing the buoyancy jump and, consequently, limiting the extension of the mixed zone (see Sec.~\ref{subsec: Entrainment_law}). This means that our correlation could be compatible with expectations from recent works about entrainment \citep{meakin&arnett2007}. \par

We investigated which are the main contributors to the overshooting efficiency uncertainty, and we found that the overshooting efficiency is mainly affected by $\MH$ and, secondarily, by the measurements of effective temperatures and distance modulus (see Sec.~\ref{subsec: Overshooting_calibration}). On the other hand, we found that our results are not significantly affected by by variations of the mixing length parameter (see Sec.~\ref{subsec: mixing_length}) or by the choice of the helium-to-metals enrichment ratio (see App.~\ref{app: Y_dY/dZ}).  \par 

A further step in this work would be to compare our findings to predictions from 3D hydrodynamical simulations \citep{blouin2023} and try to define the overshooting efficiency as a function of metallicity, in order to reduce the number of free parameters in stellar models. Finally, in addition to the indirect tests of envelope mixing presented here, high-precision asteroseismic observations across a broad range of globular clusters (GCs), such as those proposed by the candidate space mission HAYDN \citep{miglio2021b}, would provide more detailed, localized inferences on the properties of convective envelopes and mixing through the study of individual oscillation modes \citep{lindsay2022}. This would offer valuable empirical insights into the physical nature of convective boundary mixing.

\begin{acknowledgements}
L.B. aknowledges financial support from MUR (Ministero dell'Università e della Ricerca) and NextGenerationEU throughout the PNRR ex D.M. 118/2023. L.B. acknowledges Nadège Lagarde, Lorenzo Martinelli, David Nataf and Walter van Rossem for their useful suggestions. 
M.T., A.Miglio, and M.M. acknowledge support from the ERC Consolidator Grant funding scheme (project ASTEROCHRONOMETRY, G.A. n. 772293 \url{http://www.asterochronometry.eu}).\\
A.Mazzi acknowledges financial support from Bologna University, ``MUR FARE Grant Duets CUP J33C21000410001''.
A.Mucciarelli acknowledges support from the project "LEGO – Reconstructing the building blocks of the Galaxy 
by chemical tagging" (P.I. A. Mucciarelli). granted by the Italian MUR through contract PRIN 2022LLP8TK\_001. S.K. is funded by the Swiss National Science Foundation through an Eccellenza Professorial Fellowship (award PCEFP2\_194638). \\
We thank the anonymous referee for giving us comments which helped us improve the paper.

\end{acknowledgements}

\bibliographystyle{aa}
\bibliography{References.bib}

\appendix

\section{Variation of the RGBb luminosity with \texorpdfstring{$Y$}{Y} and \texorpdfstring{$\dYdZ$}{dY/dZ}}
\label{app: Y_dY/dZ}

As mentioned in Sec.~\ref{subsec: Data_Clusters}, we chose GCs such that the difference in $\ch{He}$ mass fraction between 1G and 2G stars was not larger than 0.02 at maximum and 0.01 on average, in order to avoid the effects of $Y$ on the RGBb location. Indeed, a higher value of $Y$ produces brighter and hotter stars \citep{fagotto1994}, with consequently a more luminous RGBb \citep{cassisi&salaris1997, salaris2006}. \par

\begin{figure}[h!]
   \centering
   \includegraphics[width=\columnwidth]{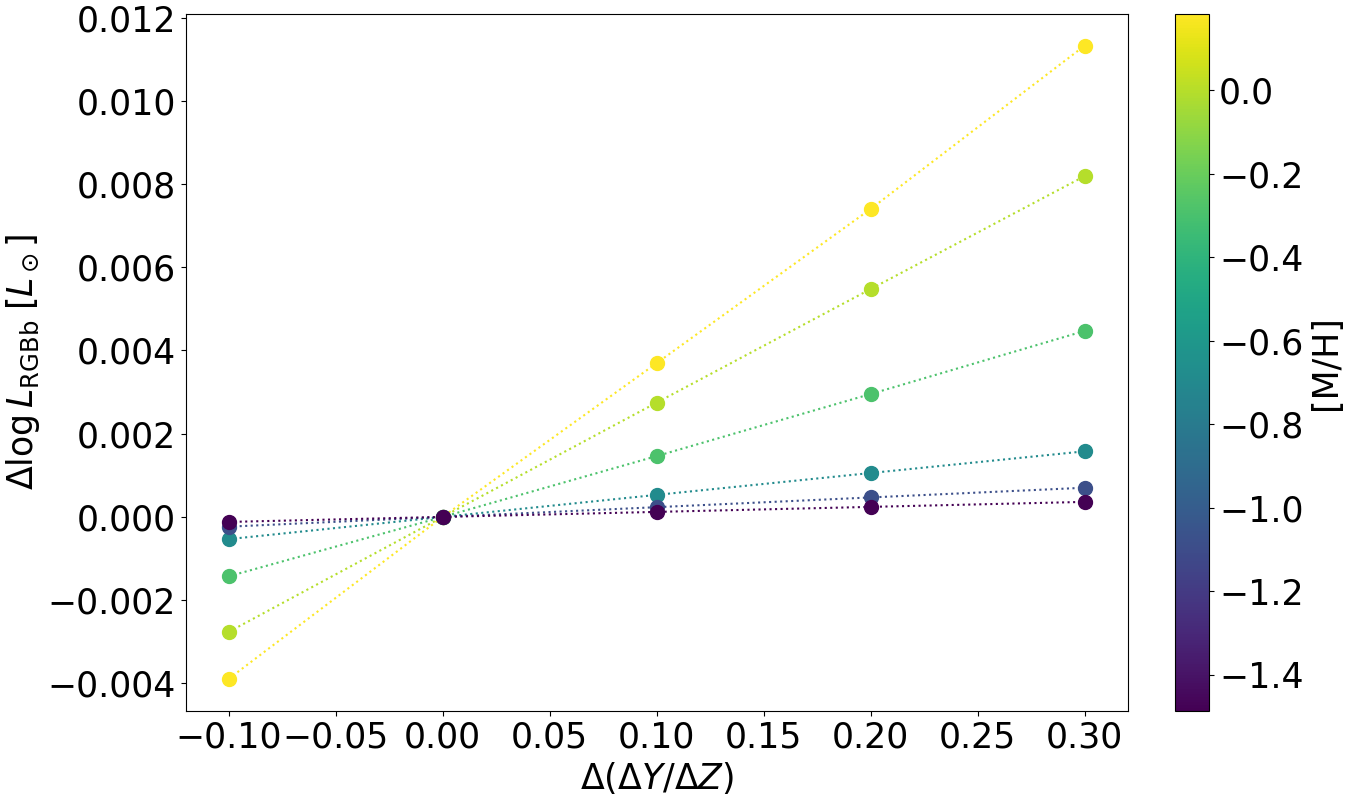} 
   
   \caption{Variation of the RGBb position in $\log L$ as a function of the variation of $\dYdZ$ for different values of $\MH$.}
    \label{fig: dYdZ}
    \end{figure}

As already mentioned in \cite{khan2018}, this fact may cause a dependence of the RGBb location on $\dYdZ$, since an $\ch{He}$ enrichment at fixed $Z$ produces higher RGBb luminosities. We thus tested how much the RGBb luminosity is affected by variations of the enrichment ratio and found that at low metallicities the effects are negligible, while at super-solar metallicities ($\MH = +0.18$), to obtain luminosity variations comparable to the errors on our measurements we need to increase $\Delta Y/ \Delta Z$ by at least 0.3 (see Fig.~\ref{fig: dYdZ}). This means that the dependence of our results on the choice of $\dYdZ$ is negligible with respect to the typical errors on our measurements in $\log L_\mathrm{RGBb}$.  
  
\FloatBarrier

\section{Script testing}
\label{appendix: Script_testing}
  
   \begin{figure}
   \centering
   \includegraphics[width=\columnwidth]{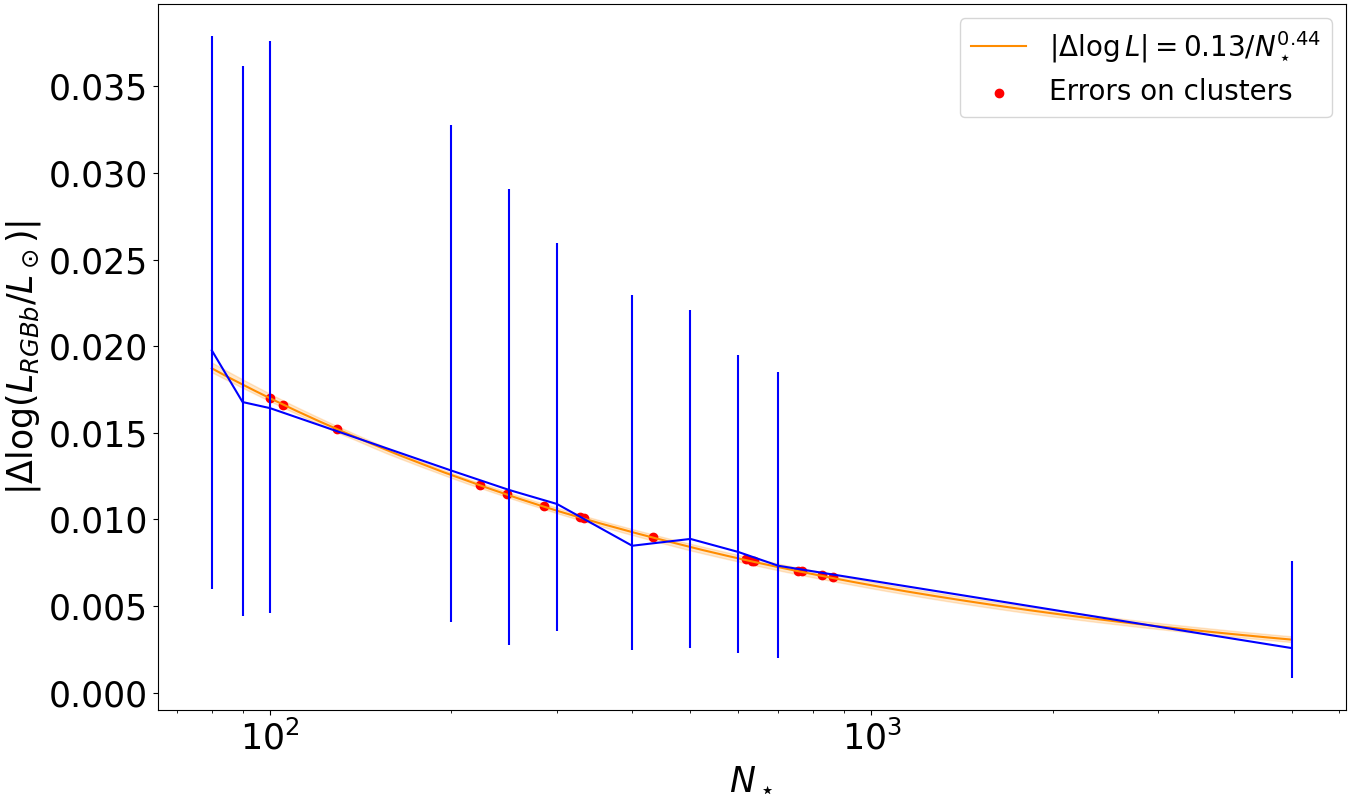} 
   \includegraphics[width=\columnwidth]{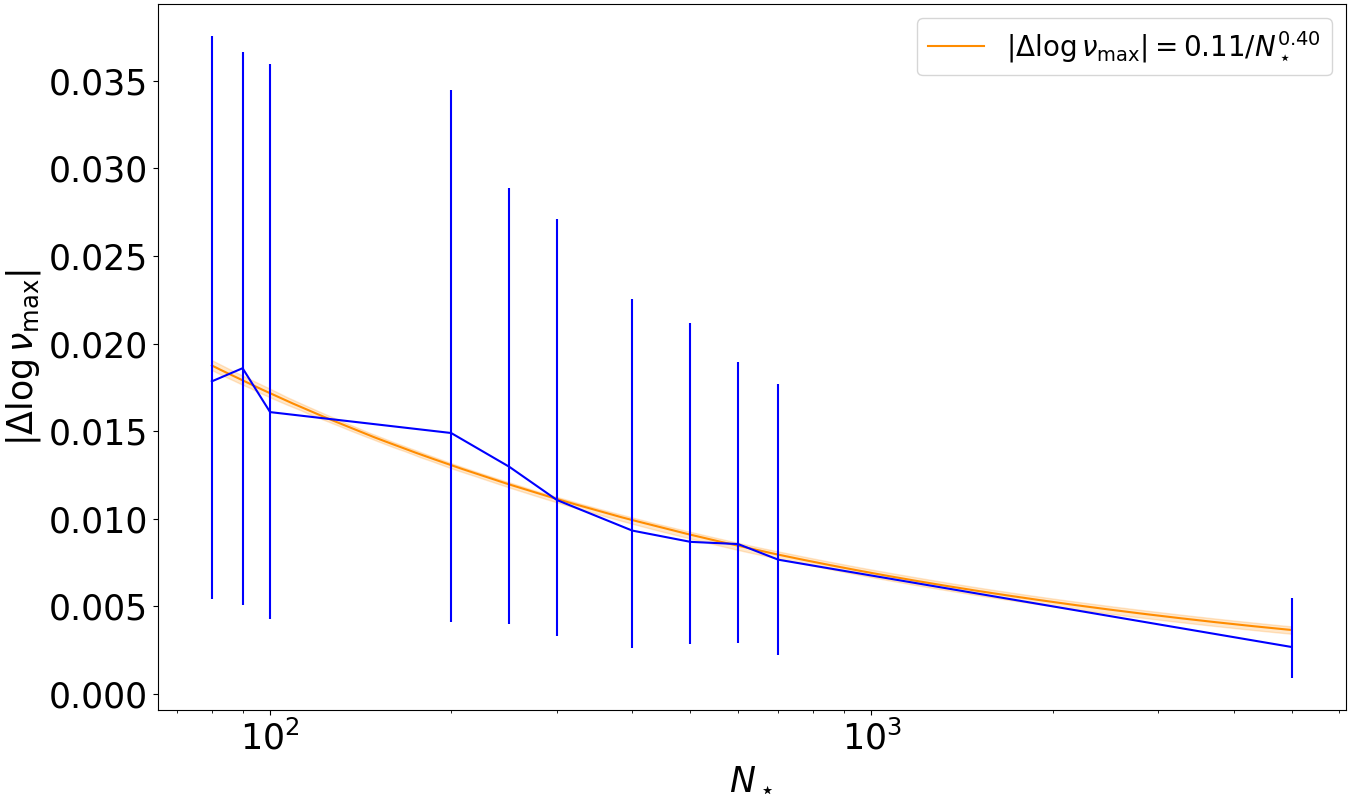} 
   \caption{Variation of the RGBb position in $\log L$ (upper panel) and $\log\numax$ (lower panel) with the number of stars in the simulation. The red points in the upper panel represent the GCs in our sample.}
              \label{fig: Script_testing}
    \end{figure} 

To test our analysis method and estimate the systematics due to the number of stars in a sample and the chosen KDE bandwidth, we performed a series of fits on synthetic populations (see Sec.~\ref{subsec: Methods_Models}) generated from the same model ($M = 0.8\; \Msun$, $\FeH = -1.50$ \si{dex}, $\aFe = +0.4$ \si{dex}, $\fov = 0.0$), varying each time the number of stars ($N_\star$) in the simulation and the bandwidth ($BW$) of the KDE, defined as
\begin{equation}
    BW = F\frac{x_\mathrm{max}-x_\mathrm{min}}{\sqrt{N_\star}},
\end{equation}
where $F$ is a multiplicative factor which we varied from 0.25 to 1.00. The fit of each $F-N_\star$ combination was performed 1000 times, to reduce as much as possible numerical effects. \par

All the results of the fits computed with $F=0.50$ (which is the value that we finally adopted for clusters and field stars) were compared with the fit of a reference simulation, generated with a large number of stars ($N_\star = 5000$) and a narrow bandwidth ($F = 0.25$): both for $\log L_\mathrm{RGBb}$ and $\log\nu_\mathrm{max, RGBb}$ we obtained that the discrepancy with the reference measurements decays hyperbolically with $N_\star$, as shown in Fig.~\ref{fig: Script_testing}. The inferred systematic error for each stellar cluster or bin of field stars was summed quadratically to the random one.

\FloatBarrier

\section{Coefficients of the Mass-Age-Metallicity scaling relation}
\label{appendix: scaling_relation}

Coefficients of the Mass-Age-Metallicity scaling relation (eq.~\ref{eq: mass-age-mh_relation}) for different values of the overshooting efficiency $\fov$.

\begin{table}[h!]
\caption{Coefficients for the scaling relation~\ref{eq: mass-age-mh_relation} obtained with different overshooting efficiencies.}            
\label{tab:Scaling_coefficients}     
\centering                          
\begin{tabular}{c c c}        
\hline\hline                
$\fov$ & $A$ & $\alpha$ \\    
\hline                       
    0.000 & $1.909 \pm 0.005$ & $-0.267 \pm 0.002$ \\
    0.025 & $1.908 \pm 0.005$ & $-0.267 \pm 0.002$ \\
    0.050 & $1.907 \pm 0.005$ & $-0.267 \pm 0.002$ \\
    0.075 & $1.906 \pm 0.005$ & $-0.267 \pm 0.002$ \\
    0.100 & $1.904 \pm 0.005$ & $-0.267 \pm 0.002$ \\
    0.125 & $1.901 \pm 0.005$ & $-0.267 \pm 0.002$ \\
\hline\hline   
$\fov$ & $B$ & $\beta$ \\    
\hline                       
    0.000 & $0.156 \pm 0.003$ & $0.57 \pm 0.02$ \\
    0.025 & $0.156 \pm 0.002$ & $0.57 \pm 0.02$ \\
    0.050 & $0.155 \pm 0.002$ & $0.57 \pm 0.02$ \\
    0.075 & $0.155 \pm 0.002$ & $0.57 \pm 0.02$ \\
    0.100 & $0.154 \pm 0.002$ & $0.57 \pm 0.02$ \\
    0.125 & $0.154 \pm 0.002$ & $0.57 \pm 0.02$ \\
\hline
\end{tabular}
\end{table}

\label{LastPage}
\end{document}